\documentstyle[fleqn,twoside,epsf]{article}

\def\to{\rightarrow}

\def\gesim{\,{\raise-3pt\hbox{$\sim$}}\!\!\!\!\!{\raise2pt\hbox{$>$}}\,}
\def\lesim{\,{\raise-3pt\hbox{$\sim$}}\!\!\!\!\!{\raise2pt\hbox{$<$}}\,}
\def\boldoverdot{\,{\raise6pt\hbox{\bf.}\!\!\!\!\>}}

\def\etal{{\it et. al.}}

\def\lcal{{\cal L}}

\def\ocal{{\cal O}}

\def\vev{vacuum expectation value}

\def\diag{\hbox{\diag}}
\def\sm{Standard Model}
\def\doubleundertext#1{
{\undertext{\vphantom{y}#1}}\par\nobreak\vskip-\the\baselineskip\vskip4pt%
\undertext{\hbox to 2in{}}}
\def\inbox#1{\vbox{\hrule\hbox{\vrule\kern5pt
     \vbox{\kern5pt#1\kern5pt}\kern5pt\vrule}\hrule}}
\def\sqr#1#2{{\vcenter{\hrule height.#2pt
      \hbox{\vrule width.#2pt height#1pt \kern#1pt
         \vrule width.#2pt}
      \hrule height.#2pt}}}
\def\square{\mathchoice\sqr56\sqr56\sqr{2.1}3\sqr{1.5}3}
\def\today{\ifcase\month\or
  January\or February\or March\or April\or May\or June\or
  July\or August\or September\or October\or November\or December\fi
  \space\number\day, \number\year}
\def\pmb#1{\setbox0=\hbox{#1}%
  \kern-.025em\copy0\kern-\wd0
  \kern.05em\copy0\kern-\wd0
  \kern-.025em\raise.0433em\box0 }
\def\up#1{^{\left( #1 \right) }}
\def\su#1{{SU(#1)}}
\def\ui{U(1)}
\def\sumprime_#1{\setbox0=\hbox{$\scriptstyle{#1}$}
  \setbox2=\hbox{$\displaystyle{\sum}$}
  \setbox4=\hbox{${}'\mathsurround=0pt$}
  \dimen0=.5\wd0 \advance\dimen0 by-.5\wd2
  \ifdim\dimen0>0pt
  \ifdim\dimen0>\wd4 \kern\wd4 \else\kern\dimen0\fi\fi
\mathop{{\sum}'}_{\kern-\wd4 #1}}
\def\d0{D\O}
\def\D0{D\O}
\def\w{$W$}
\def\W{$W$}
\def\z{$Z$}

\def\wg{$W\gamma$}
\def\ww{$WW$}

\def\wz{$WZ$}
\def\wwg{$WW\gamma$}

\def\wwz{$WWZ$}
\def\zg{$Z\gamma$}
\def\zzg{$ZZ\gamma$}
\def\zgg{$Z\gamma\gamma$}
\def\pt{$p_T$}
\def\et{$E_T$}
\def\etmisv {\mbox{${\hbox{${\vec E}$\kern-0.6em\lower-.1ex\hbox{/}}}_T$}}
\def\etmis  {\mbox{${\hbox{$E$\kern-0.6em\lower-.1ex\hbox{/}}}_T$}}

\def\ifmath#1{\relax\ifmmode #1\else $#1$}%
\def\TeV{\ifmmode {\mathrm{ Te\kern -0.1em V}}\else
                   \textrm{Te\kern -0.1em V}\fi}%
\def\GeV{\ifmmode {\mathrm{ Ge\kern -0.1em V}}\else
                   \textrm{Ge\kern -0.1em V}\fi}%
\def\MeV{\ifmmode {\mathrm{ Me\kern -0.1em V}}\else
                   \textrm{Me\kern -0.1em V}\fi}%
\def\GeVcc{\ifmmode {\mathrm{ \GeV/c^2}}\else
                   \textrm{Ge\kern -0.1em V/c$^2$}\fi}%
\def\MeVcc{\ifmmode {\mathrm{ \MeV/c^2}}\else
                   \textrm{Me\kern -0.1em V/c$^2$}\fi}%

\def\ppbar              {\mbox{$p\overline{p}$}}

\def\etal{{\it et al.}}

\def\geant{{\sc geant}}

\def\vecbos{{\sc vecbos}}
\def\herwig{{\sc herwig}}
\def\qwe{\alpha}

\def\sm{standard model}
\newcommand{\tab}{\hspace*{0.5in}}

\begin{document}
\pagestyle{empty}

\vskip -4cm 
\rightline{UCR/{D\O}/98-01}
\rightline{UCRHEP-T227}
\vskip 1cm

\noindent
{\huge Study of Trilinear Gauge Boson\\
Couplings at the Tevatron Collider}
\bigskip

\noindent
{\large {\it John Ellison and Jos\'{e} Wudka}}

\medskip
\noindent
Department of Physics, 
University of California,\\
Riverside, California 92521

\bigskip
\noindent
{\sc key words}:\quad  Tevatron, CDF, \d0, standard model, gauge boson
couplings

\bigskip
\hrule
\bigskip

\begin{abstract}
We review studies of the trilinear gauge boson
couplings at the Tevatron proton-antiproton collider from data
collected by the CDF and \d0\ collaborations during the period
1992--1996. The gauge boson couplings are a fundamental prediction of the
standard model, resulting from the non-Abelian nature of the theory.
Therefore, experimental tests of the couplings are of foremost importance.
We introduce the experimental results by reviewing the effective
Lagrangian formalism, the indirect constraints on the couplings from
low-energy experiments, and the expected values of the couplings in 
theories beyond the standard model. Finally, we consider the prospects
for future measurements.

\end{abstract}
\bigskip
\hrule
\bigskip

\begin{center}
{\it (To be published in  Annual Review of Nuclear
and Particle Science.)}\\
\end{center}

\newpage

\pagestyle{myheadings}

\setcounter{tocdepth}{2}
\bigskip
{\footnotesize
\tableofcontents
}
\bigskip


\section{\rm \large VECTOR BOSON SELF-INTERACTIONS AND\\
EFFECTIVE COUPLINGS}

The hallmark of the \sm\ (SM) is gauge invariance under the group 
\su3~$\times$~\su2~$\times$~\ui~\cite{sm.general}. The consequences of this
symmetry are manifold,
ranging from universal coupling of matter fields to the prediction of
the vector boson self-couplings. This symmetry lies at the heart of the
model, and its consequences should be investigated as deeply as
possible.

Among the many probes devised to study the gauge symmetry of
the SM, the experiments designed to investigate the gauge boson 
self-couplings have received much attention~\cite{tbc.general}. 
This interest is generated by 
the fact that these interactions are intimately
related to the gauge group of the model, and 
a deviation from the SM would provide important information
about the kind of new physics beyond the SM. 
The possible trilinear couplings involving the 
electroweak gauge bosons $W^\pm$, $Z^0$, and $\gamma$
are the \wwg, \wwz, \zzg, \zgg, and $ZZZ$ couplings. Only the first two
are allowed in the \sm\ at tree level.
These vertices have
been directly probed at the Tevatron by the D\O\ and CDF experiments, based
on 100 pb$^{-1}$ of data collected in Run I at the Tevatron 
proton-antiproton collider during 1992--1996. The
measurements provide an important confirmation of the gauge
structure of the SM. 

In this article we consider the physics of triple vector boson
couplings at the Tevatron collider.
We investigate the sensitivity of current
experiments to the SM predictions and to possible deviations 
generated by 
new physics. We assume that the physics
responsible for these deviations  is not directly
observed and can be probed only though virtual effects. Hence it can
be studied using an effective Lagrangian
approach~\cite{leff.general}. 
This formalism provides a simple parametrization
of all heavy particle effects at low energies in terms of
a set of unknown constants, the magnitudes of which can be bounded using 
experimental data and estimated for various classes of models. From
these results, information about any new interactions can be extracted.
With this information, 
dedicated experiments can be designed to probe these new interactions 
directly.
This approach is consistent with the gauge structure of the SM
and is both model and process independent.

\subsection{\it Effective Lagrangians and Form Factors}
\label{sec: eff.l}

Despite the successes of the SM, it is widely believed that this
model represents only the low-energy limit 
of a more fundamental theory [see, for example, Weinberg~\cite{new}].
If so, the
relevant question is then whether current data provides any
guidelines as to what kind of new physics underlies the SM. 
Theoretical constraints have led to the development of specific 
models~\cite{models}, yet to date there is no experimental evidence of any 
non-SM physics~\cite{pdg}, and so the only unavoidable requirement of a
model is that it reproduce the SM results to within the
experimental precision.

It is therefore reasonable to use a model-independent
effective Lagrangian parametrization of non-SM physics.
We follow this
route here and avoid  choosing any one specific theory (except as an illustration). Our only assumption is that
the new physics is not observed directly---that is, that the scale of new
physics, which we will denote by $ \Lambda $, lies above the energy
available to the experiments. The
approach fails if this condition is violated.

This approach implicitly requires
complete knowledge of the low-energy particle spectrum, on which the
results depend strongly. We assume that the only 
light excitations correspond to the SM fermions,
gauge bosons, and possibly scalars; concerning the latter we will
present results for the case where there is a single physical scalar
(the usual SM Higgs boson) and for the case where there are no
light scalars at all---the so-called chiral case~\cite{chiral.leff}. 
The more complicated
possibilities of an extended light gauge group~\cite{frere.et.al.} or
of a scalar sector containing more than one light
multiplet~\cite{perez.et.al.} can be studied along the same lines
but will not be considered here.

\def\fh{\Phi}
\def\fl{\phi}

We consider situations in which there are two types of fields, denoted
collectively by $ \fh $ and $ \fl $, whose scales lie respectively
above and significantly below a scale $ \Lambda $. We assume that the fields
$\fh $ are not observed directly, but affect currently measured 
observables through 
virtual effects that can be summarized by a series of
effective vertices containing only internal heavy lines and external
light lines. The nonlocal and nonlinear functional of $ \fl $ that 
generates these effective vertices is called the effective action $
S_{\rm eff} [ \fl ] $. 

At low energies (i.e. below $ \Lambda $) all processes 
can be calculated using $ S_{\rm eff}
[ \fl] + S_{\rm light} [\fl] $ where $ S_{\rm light} $ contains all
interactions among the light excitations present in the original theory;
for the case at hand it corresponds to the SM action.
The effective action must be invariant under the gauge
transformations obeyed by $ S_{\rm light} $,
otherwise
there is no natural way of defining the gauge symmetry of the
light theory~\cite{veltman.ir.uv.}.

The effective action $ S_{\rm eff} [ \fl] $ contains the scale $
\Lambda $ as a parameter. For the situations under consideration, all
energies and light masses will be significantly below $ \Lambda $, and
hence
an expansion in powers of $ 1/ \Lambda $ of the
effective vertices constituting $ S_{\rm eff} $ is appropriate.
The terms in this expansion are all local operators.
For the case where
the underlying theory decouples, all terms with nonnegative powers of $
\Lambda $ renormalize the parameters of 
$ S_{\rm light} $~\cite{leff.general,arzt.et.al.mom}. Where the
underlying theory does not decouple, this expansion corresponds to a
derivative expansion~\cite{chiral.leff}. In either case we can write
\begin{equation}
S_{\rm eff} [ \phi] = \int d^4x \lcal_{\rm eff} \qquad
\lcal_{\rm eff} [ \phi] =  \sum_n {1 \over \Lambda^n } \sum_i
\qwe_i\up n \ocal_i\up n
\label{leff.generic},
\end{equation}
where $\lcal_{\rm eff} [ \phi]$ is the effective Lagrangian and 
the operators $ \ocal_i\up n$ have dimension [mass]$^{n-4}$, are local
functions of the light fields, and obey the same gauge symmetries as $
S_{\rm light}$. The coefficients $\qwe$ are obtained from the parameters in the
original theory. In general, all possible operators $ \ocal $ allowed by
the local symmetries will be induced\footnote{For some particular underlying
theories, however, some operators might be absent as a result of some
additional symmetries not apparent in $ S_{\rm light} $.}, and because
of this the coefficients $\qwe$ in the above expansion parametrize all
possible effects
at low energies. These parameters can be estimated by requiring
consistency of the underlying theory (see Section
{\ref{sec: coefficient.estimates}}).

In practical applications, the infinite summation over $n$ in 
Equation~\ref{leff.generic} is cut at some finite 
value $n_o$. This approximation is
appropriate since, by assumption, all external momenta to a given process
lie significantly below $ \Lambda $. In this case, we can use the
coefficient estimates together with the generic form of the operators
appearing at the next order to estimate the error made in eliminating the
terms with $ n > n_o $.

It must be noted that the choice of operators is not universal.
If the difference between two operators $ \ocal_1 $ and $ \ocal_2 $ 
vanishes when the light equations of motion are used,
then the corresponding coefficients $ \qwe_1 $ and $ \qwe_2 $ appear in all
observables only in the combination $ \qwe_1+\qwe_2 $~\cite{eom}.
This fact can be used to choose an irreducible operator basis~\cite{bw},
but the bases differ from one publication to another.
The underlying interactions may of course
generate all operators whether they are redundant or not.

The effective Lagrangian parametrization is completely general and
consistent, but it will fail at energies
close to $ \Lambda $, for in this case all terms in the expansion in $n$
become equally significant. For the same reason it makes no sense to
test the unitarity of the theory at arbitrarily large energies.

\subsubsection{\footnotesize TRIPLE GAUGE BOSON VERTICES}
\label{sec:lagrangians}

For our discussion, the relevant terms in $ \lcal_{\rm eff} $ are
those that produce vertices with three or four gauge bosons. Operators
containing fermions do not contribute to these vertices. In contrast,
operators containing scalars may contribute, since upon 
spontaneous breaking of the
gauge symmetry such scalar fields may acquire a \vev.

The notation used for the triple gauge vertices involving two
$W$ bosons is~\cite{hagiwara.peccei}
\begin{eqnarray}
\lcal_{WWV} / g_{WWV} & = & 
i g_1^V   \left( W_{\mu\nu}^\dagger W^\mu V^\nu 
                - W_\mu^\dagger V_\nu W^{\mu\nu} \right) 
+ i \kappa_V W_\mu^\dagger W_\nu V^{\mu \nu } \cr
& + & i { \lambda_V \over m_W^2} W_{\lambda \mu }^\dagger W^\mu{}_\nu V^{\nu \lambda }
- g_4^V W_\mu^\dagger W_\nu \left( \partial^\mu V^\nu +
\partial^\nu V^\mu \right) \cr
& + & g_5^V \epsilon^{\mu\nu\lambda \rho} \left( 
W_\mu^\dagger \partial_\lambda W_\nu -
\partial_\lambda W_\mu^\dagger \, W_\nu 
\right) V_\rho \cr
& + & i \tilde \kappa_V W_\mu^\dagger W_\nu \tilde V^{\mu \nu }
+ i { \tilde \lambda_V \over m_W^2} W_{\lambda \mu }^\dagger 
W^\mu{}_\nu \tilde V^{\nu \lambda } 
\label{wwv},
\end{eqnarray} 
where 
$W$ denotes the $W$ boson field,
$V = Z$ or $\gamma $,
$ V_{\mu\nu} = \partial_\mu V_\nu - \partial_\nu V_\mu $ (and similarly for $ W_{\mu \nu} $),
$ \tilde V_{\mu \nu} = {1\over2} \epsilon_{\mu\nu\alpha\beta}
V^{\alpha \beta } $, $ g_{WW \gamma}=-e, \; g_{WWZ}=-e \cot 
\theta_W $, and $\theta_W$ denotes the weak mixing angle.
In the SM at tree level the values of the couplings are 
$\kappa_V = g_1^V = 1$, and 
$\lambda_V = \tilde \lambda_V = \tilde \kappa_V = g_4^V = g_5^V = 0$. 
We define $\Delta\kappa_V \equiv
\kappa_V -1$ and $\Delta g_1^V \equiv g_1^V - 1$, which are both zero
at tree level in the SM. 
The couplings $g_4^V$, $\tilde \kappa_V$, and $\tilde \lambda_V$
violate $CP$ invariance, while all other couplings are $CP$ conserving.
Equation \ref{wwv} is obtained from Equation \ref{leff.generic} by replacing all
scalar fields with their vacuum expectation values and selecting all
terms with three gauge bosons, two of which are $W$s.

Similarly, the triple gauge boson vertices involving one $Z$ boson and
one photon with both on shell are given by
\begin{eqnarray}
\lcal_{Z \gamma V } &=& - i e \Biggl[ 
\left( h_1^V        F^{ \mu \nu }  +
h_3^V \tilde F^{ \mu \nu } \right) Z_\mu 
{ \left( \square + m_V^2 \right) \over m_Z^2}  V_\nu
\cr &&
+ \left( h_2^V F^{ \mu \nu } +
 h_4^V  \tilde F^{ \mu \nu } \right) Z^\alpha 
{ \left( \square + m_V^2 \right)  \over m_Z^4}\partial_\alpha \partial_\mu V_\nu
\Biggr],
\label{zgv}
\end{eqnarray}
where $ F^{\mu \nu} $ denotes the photon field strength (note that
$V$ is not necessarily on shell).
The couplings $h_1^V$ and $h_2^V$ violate $CP$ invariance, while
$h_3^V$ and $h_4^V$ are $CP$ conserving.
There is a corresponding set of
vertices describing the interactions of two on-shell $Z$ bosons with a
$Z$ or photon, but these parameters are not accessible at
current experimental energies and luminosities.
At tree level, the SM values for the coefficients $ h_i^V $ are zero.
Henceforth we use the terms effective coupling and anomalous coupling for the coefficients in Equations~\ref{wwv} and
\ref{zgv}.

The coefficients in Equations~\ref{wwv} and \ref{zgv} have the following 
relation to physical quantities:
\begin{equation}
\begin{tabular}{ll}
$\mu_W = { e \over 2 m_W } ( 1 + \kappa_\gamma + \lambda_\gamma)$
&
$Q^e_W = - { e \over m_W^2 } ( \kappa_\gamma - \lambda_\gamma )$ 
\\
$d_W = { e \over 2 m_W } ( \tilde \kappa_\gamma + \tilde \lambda_\gamma )$
&
$Q^m_W = - { e \over m_W^2 } ( \tilde \kappa_\gamma - \tilde \lambda_\gamma ) $
\\
$\mu_Z = { - e \over \sqrt{2} \, m_Z} { E_\gamma^2 \over m_Z^2} \left( h_1^Z - h_2^Z \right)$ 
&
$Q^e_Z = { 2 \sqrt{10} \, e \over m_Z^2} h_1^Z $
\\
$d_Z = { - e \over \sqrt{2} \, m_Z} { E_\gamma^2 \over m_Z^2} \left( h_3^Z - h_4^Z \right)$ 
&
$Q^m_Z = { 2 \sqrt{10} \, e \over m_Z^2} h_3^Z $,
\\
\end{tabular}
\end{equation}
where $\mu $ and $d$ denote the magnetic and electric dipole moments and 
$ Q^m $, $ Q^e $ the corresponding quadrupole moments of the $W$
and $Z$ bosons.  
For the $W$ these are static moments and for the
$Z$ these refer to the transition moments where $ E_\gamma $ is
the photon energy~\cite{renard-moments}.


The notation used in Equations \ref{wwv} and \ref{zgv} 
is far from universal. The LEP groups
have proposed a different parametrization~\cite{future.39} in terms of
the operator
coefficients $ \qwe $ for a particular choice of operator basis and
with $ \Lambda $ replaced by $ m_W $. Specifically,
\begin{eqnarray}
\lcal =
&  i g' { \qwe_{ B \phi } \over m_W^2} 
\left( D^\mu \phi \right)^\dagger \left( D^\nu \phi \right) B_{\mu\nu} +
    i g { \qwe_{ W \phi } \over m_W^2} 
\left( D^\mu \phi \right)^\dagger \sigma_I \left( D^\nu \phi \right) W^I_{\mu\nu} \cr 
&+ g { \qwe_W \over 6 m_W^2 } \epsilon_{IJK} 
W^I{}^\mu_\nu W^J{}^\nu_\rho W^K{}^\rho_\mu +
g { \qwe_{\tilde W} \over 6 m_W^2 } \epsilon_{IJK} 
W^I{}^\mu_\nu W^J{}^\nu_\rho \tilde W^K{}^\rho_\mu \cr
&+  i g' { \qwe_{ \tilde B \phi } \over m_W^2} 
\left( D^\mu \phi \right)^\dagger \left( D^\nu \phi \right) 
\tilde B_{\mu\nu} +
i g { \qwe_{ \tilde W \phi } \over m_W^2} 
\left( D^\mu \phi \right)^\dagger \sigma_I \left( D^\nu \phi \right) 
\tilde W^I_{\mu\nu},
\label{lep.leff}
\end{eqnarray}
where $\phi$ denotes the SM scalar doublet, $B_{\mu \nu} $ the field
strength for the U(1) gauge field, and $ W_{\mu\nu}^I$ the non-Abelian
SU(2) field strength; $g$ and $g^\prime$ are the corresponding gauge 
coupling constants
and $D$ denotes the covariant 
derivative\footnote{The choice of effective operators is also
not universal (see Section~{\ref{sec: eff.l}}), even in the number of
parameters. For example, in Ref.~\cite{bw} only four operators of dimension six contribute to Equation~\ref{wwv}.}. From Equation \ref{lep.leff} one finds
\begin{eqnarray}
\Delta g_1^Z = { \qwe_{W \phi} \over c_W^2} &\qquad&
\lambda_\gamma = \lambda_Z=\qwe_W \cr
\Delta \kappa_\gamma = \qwe_{W \phi } + \qwe_{ B \phi} &\qquad&
\Delta \kappa_Z = \qwe_{W \phi } - { s_W^2 \over c_W^2} \qwe_{ B \phi},
\label{hisz.scenario} 
\end{eqnarray}
where $s_W $ and $ c_W $ denote the sine and cosine of the weak mixing
angle. This parametrization includes only dimension six 
operators; to this order the remaining couplings in Equation~\ref{wwv} are
zero.
The advantage of this approach
is that the original expressions are manifestly gauge invariant.
The disadvantages are, first, the
neglect of operators of dimension eight 
(leading to, for example, $ \lambda_Z \not= \lambda_\gamma $), 
which is justified only when $ \Lambda \gesim 3 $~TeV
(see Section {\ref{sec: coefficient.estimates}}), and second,  
the use of $m_W$ instead of $\Lambda$ to set the scale of the operators, 
which buries the dependence on the scale of new physics inside the
coefficients. This is not inconsistent, but it
obfuscates the virtues of the effective Lagrangian approach. 

It is worth pointing out that Equation~\ref{hisz.scenario} expresses
$ \Delta g_i^Z,\; \lambda_\gamma, \;  \lambda_Z,
\; \Delta \kappa_\gamma$ and $\Delta \kappa_Z$
in terms of three parameters. These relations are not
a consequence of gauge invariance but result solely from ignoring
operators of dimension eight in the linear case. [In the chiral case
these relations do not hold; in particular 
$ \lambda_\gamma = \lambda_Z = 0$~\cite{appelquist.wu}.]
If in addition it is assumed (for simplicity only) that $ \qwe_{ W \phi
} = \qwe_{ B \phi }$, the so-called HISZ 
scenario~\cite{hisz.ref}, then only two 
parameters determine the CP-conserving couplings of Equation~\ref{wwv}.

In contrast to the LEP parametrization, 
the parameterization presented in Equations \ref{wwv} and \ref{zgv}
is completely genera. (The contributions from operators
of arbitrarily large dimension will take the same form when
the coefficients are replaced by appropriate functions of the
Lorentz invariant Mandelstam parameters.) The disadvantage
of Equations \ref{wwv} and \ref{zgv} is that the expressions are not
manifestly gauge invariant.
In this paper we choose to sacrifice explicit gauge invariant
expressions in favor of the greater generality of
Equations \ref{wwv} and \ref{zgv}.\footnote{An approach that would
maintain both generality and gauge invariance would necessitate
the itemization of all gauge invariant operators of dimension eight, 
which has not been done.}

\subsubsection{\footnotesize COEFFICIENT ESTIMATES}
\label{sec: coefficient.estimates}

One of the advantages of the effective Lagrangian
formulation is that one can
obtain reliable bounds on the coefficients $ \qwe $. These bounds are
obtained from general considerations and are verified in all models
where calculations have been performed. In this subsection we distinguish two cases: that in which the underlying theory is weakly
coupled and for which there are light scalars, and that for which the
symmetry-breaking mechanism is generated by a new type of strong
interaction (such as technicolor). The first case we label the
linear case, the second the chiral case.

Within the linear case, the underlying physics is 
expected to be weakly coupled~\cite{arzt.et.al.loops,wudka.rev},
and the magnitude of the coefficient of a given
operator is determined by whether it is generated at tree level
or via loops by the heavy physics. Loop-generated operators are
subdominant since their coefficents are
suppressed by a factor $ \sim 1/(4
\pi)^2 $ relative to the coefficients of a tree level--generated
operator.

In the linear case,
the terms in Equation~\ref{wwv}
proportional to $ \kappa_V $, $ \tilde \kappa_V $,
$ \lambda_V $, and $ \tilde \lambda_V $ are generated by dimension six
operators; all seven terms are generated by dimension eight
operators~\cite{hisz.ref,wudka.rev}. Similarly,
the terms proportional 
to $ h^V_1 $ and  $ h^V_3 $ 
are generated by dimension
eight operators, while  the terms
containing $ h^V_2 $ and  $ h^V_4 $ 
are generated by operators of dimension ten.
The dimension six operators are necessarily loop
generated~\cite{arzt.et.al.loops}, while the relevant
operators of dimension eight or ten 
can be generated at tree level by
the heavy dynamics.\footnote{This means that there are
certain kinds of heavy physics that can generate these operators
at tree level. There is no guarantee that such new interactions
are allowed by all existing data, or are the ones realized 
in nature.}
It is also important to note that the $W$s are
gauge bosons and will necessarily couple with strength $g$. Collecting
these results we  estimate (where $v$ denotes the SM \vev\, $\simeq
246$~GeV)
\begin{eqnarray}
&&
\left| \Delta \kappa_V \right| , \; 
\left| \tilde \kappa_V \right|  
\lesim \max\left\{ \left( { m_W \over 4 \pi \Lambda } \right)^2 , \left( {
m_W v \over \Lambda^2 } \right)^2 \right\} \cr
&&
\left| \lambda_V \right| , \; 
\left| \tilde \lambda_V \right| 
\lesim \max\left\{ \left( {g m_W \over 4 \pi \Lambda } \right)^2 , \left( {
m_W   \over \Lambda  } \right)^4 \right\} \cr
&&\left| \Delta g_1^V \right|, \; \left| g_{4,5}^V \right| 
\lesim { m_W^2 v^2 \over \Lambda^4 }
\qquad \left|h^V_{1,3}\right| \lesim { m_Z^4 \over \Lambda^4} 
\qquad 
\left| h^V_{2,4} \right| \lesim { m_Z^6 \over \Lambda^6}. 
\label{estimates.weak}
\end{eqnarray}
Higher dimensional operators generate corrections smaller
by factors of $( v/\Lambda )^2 $ or $ (E/\Lambda)^2 $ 
where $E$ is a typical energy in the vertex. These values are very small
within the range of applicability of the effective Lagrangian formalism.

For the chiral case, a different approach must be followed because the
absence of light scalars requires the presence of a large coupling constant. This is apparent because the chiral case can be obtained
by considering the SM in the case where the Higgs mass $m_H$ is much
larger than the Fermi scale, and the only way of generating a large $ m_H$
while keeping the Fermi constant $G_F$ fixed is to require the
scalar self-coupling to be $\gg 1$, whence the scalar sector is
strongly coupled. 

The coefficients $\qwe$ in the chiral case can be estimated using 
naive dimensional analysis~\cite{nda}. The basic idea is that the effective
Lagrangian at low energies, though strongly coupled, must be a 
consistent theory, i.e. the radiative corrections
obtained from it must not overwhelm the tree-level 
contributions. [Failure of this condition indicates that the
fields in the theory do not correspond to the low-energy degrees of
freedom~\cite{polchinski}.]
In the chiral case the
scale of new physics is approximately
\begin{equation}
\Lambda \sim 4 \pi v \sim 3~{\rm TeV}; \quad \hbox{(chiral case)},
\end{equation}
where $v$ denotes the SM \vev\ $ \simeq 246$~GeV. The operators
in this case are classified by their number of derivatives. Those contributing to $ \Delta \kappa_V, \;
\tilde \kappa_V, \; g_{1,4,5}^V$ contain four derivatives; those 
contributing to $ \lambda_V , \; \tilde\lambda_V $,
and $ h_{1,3}^V$ contain six derivatives; and those contributing to
$ h_{2,4}^V$ contain eight derivatives. This leads to the following estimates:
\begin{eqnarray}
\left| \Delta\kappa_V \right|, \; 
\left| \tilde \kappa_V \right|, \;
\left| \Delta g_1^V \right|, \;
\left| g_{4,5}^V \right|
& \sim & {1 \over ( 4 \pi)^2 } \simeq 0.006 \cr
\left| \lambda_V \right|,  \; \left| \tilde \lambda_V \right|, \; \left| h_{1,3}^V \right|
& \sim & {g^2 \over (4 \pi)^4 } \simeq 2 \times 10^{-5} \cr
\left| h_{2,4}^V \right |
& \sim & {g^4 \over (4 \pi)^6 } \simeq 5 \times 10^{-8}. 
\label{estimates.nda}
\end{eqnarray}
Higher dimensional operators generate corrections of order
$(E/\Lambda)^2 $ to these estimates, where $E$ is a typical energy
in  the vertex. This might suggest the possibility that at sufficiently
high energies these vertices play a dominant role. However, this is not
the case. For this to occur, we must have $ E \gesim \Lambda $,
which lies well beyond the applicability of the effective Lagrangian
parametrization. Therefore, the estimates in Equation~\ref{estimates.nda}
do provide
the theoretical upper bounds on the corresponding coefficients.
If the new physics is within the reach of the collider, then these
estimates are invalid and the whole formalism breaks down (for an example
see~\cite{frere.et.al.}.)

\subsubsection{\footnotesize FORM FACTORS}
\label{sec: form.factors}

As emphasized above, the effective interactions described in
Equations~\ref{wwv} and \ref{zgv} may not be applied at energies approaching $
\Lambda $. In this regime, all terms in Equation~\ref{leff.generic}
become equally important and must be included. If only a finite number
of terms is retained and the model is blindly applied at sufficiently 
large energies, it will exhibit serious pathologies, such as lack
of unitarity. For example, Feynman diagrams
containing vertices proportional to $ \lambda_V $ will generate
unitarity violations  at
energies $ \gesim m_W/ \sqrt{\lambda_V}$.  Using the estimates
given in Section~{\ref{sec: coefficient.estimates}}, these can be  
significantly (and often astronomically; see Equation~\ref{estimates.nda})
above $ \Lambda$.

Any tractable extension of the effective Lagrangian method to energies
at or above $ \Lambda$ requires a unitarization procedure. This can be
achieved by modifying the particle spectrum or by replacing the effective
coefficients with appropriate form factors~\cite{unit.form.factors}.
The procedure is model dependent and
in this sense deviates from the philosophy used in studying
non-SM effects using an effective Lagrangian. Other
caveats associated with the form factor approach are discussed below.

As an example we consider the reaction
$ W^+ W^- \rightarrow W^+ W^- $, which receives contributions from 
$s$-channel $Z$ and photon exchanges and depends on the vertex
in Equation \ref{wwv}. The cross section
violates tree-level unitarity whenever the center-of-mass (CM) energy
is large enough. This can be avoided by replacing any coefficient $ \qwe
$ in Equation~\ref{wwv} according to
\begin{equation}
\qwe \rightarrow { \qwe_0 \over (1 + {\hat s} / \Lambda_{FF}^2 )^n } ,
\label{form.factor}
\end{equation}
where $\sqrt{\hat s}$ is the CM energy of the scattering
process and the exponent $n$ is chosen 
to insure unitarity.\footnote{In the experimental results we discuss, 
the form factor used is that of Equation~\ref{form.factor}.
The value $n = 2$ is used for the \wwg\ and \wwz\ effective
couplings. For the \zzg\ and \zgg\ couplings, the values used are
$n=3$ for $h^V_{1,3}$ and $n=4$ for $h^V_{2,4}$. These choices
insure that unitarity is satisfied and that all couplings have the same
high-energy behavior.}
Of course, this is not the only possible choice. One alternative expression would be  
$ \qwe_0 / \left[ (1 - {\hat s} /
\Lambda_{FF}^2 )^2 + \Gamma^2 / \Lambda_{FF}^2 \right]^n $,
which has the
disadvantage of depending on a new parameter $ \Gamma $, but has the
advantage of an obvious physical interpretation as the contribution of
a resonance of width $\Gamma$ and mass $\Lambda_{FF}$.
It must also be noted that in gauge theories, individual Feynman diagrams
might violate unitarity and only the sum of all contributions is
guaranteed to behave correctly at large energies. Imagine, for example,
the presence of a new particle that modifies the $WWZ$ vertex. This
particle should then carry an \su2\ charge and will modify the $WWWW$
vertex as well as the $W$ propagator. Only the sum of all these
contributions will provide a unitary cross section (as verified by
explicit calculation). Thus it is not the contributions 
from the $WWV$ vertex alone that must
satisfy unitarity but a combination of these contributions with those generated
by a quartic vertex
$WWWW$ and a modification to the kinetic energy $W \square^2 W $~\cite{fer}. 

To summarize,
replacing the parameters in Equations~\ref{wwv} and \ref{zgv} by form
factors is a viable way
of insuring unitarity. Granted, this approach is 
model dependent, and for realistic values of the parameters it
is unnecessary because unitarity violations will occur only for very
large values of the CM energy. We note, however, that
many of the experimental
results and sensitivity estimates are given in terms of the parameters
of some form factors (such as $ \lambda_0$, $ \Lambda_{FF} $,
and $n$ in Equation~\ref{form.factor}).
In these analyses, the quantity
$ 1 + {\hat s}/  \Lambda^2_{FF} $ is greater than one, and hence any limit
obtained for a coefficient $ \qwe_0 $ in Equation~\ref{form.factor}
provides an upper bound on the sensitivity to the corresponding 
parameter $ \qwe $.

\subsection{\it Indirect Constraints on Effective Couplings}

Several precision measurements would be affected by the
presence of nonstandard values in Equations \ref{wwv} and \ref{zgv}. 
The most significant are the oblique parameters, the anomalous magnetic moment of the muon, 
the electron dipole moment, and the $ b \rightarrow s \gamma $ decay rate. 

Before
itemizing the existing constraints, it is pertinent to issue a general
warning concerning these types of constraints. By their very nature,
precision measurements are sensitive to several vertices, any or all of which
may be modified by the new interactions. Consequently, the experimental
data cannot be unambiguously
translated directly into bounds on the
coefficients of Equations~\ref{wwv} and \ref{zgv}. (Moreover, the magnitudes for some new physics
contributions might easily overwhelm those generated by 
Equations \ref{wwv} and \ref{zgv}; see 
Section {\ref{sec: coefficient.estimates}}.) 
Hence, the bounds obtained should be taken only as rough
estimates. With these caveats the existing bounds are presented in 
Table \ref{tab:table1}. In addition to these results, we note the 
bound~\cite{bound.60} $-8.6 < g_5^Z < 4.1 $ obtained from 
$ B \rightarrow X_S \nu \bar \nu $ (assuming $ g_1^Z =1)
$. We 
are not aware of any estimation of the bounds on $ h_{1,2,3,4}^V
$ generated by precision measurements.

Certain observables (such as the gauge boson masses) receive radiative corrections from the effective vertices that are
proportional to a positive power of $ \Lambda $. Such
terms are renormalization artifacts and are not 
observable~\cite{arzt.et.al.mom}, and therefore the
bounds derived from them are not reliable.
We also note that in the linear case, when operators of dimension six
dominate
($ \Lambda > 4 \pi v \sim 3$~TeV), the constraints on $ \Delta
\kappa_\gamma $ also apply to $ \Delta \kappa_Z$ 
(see {Section~\ref{sec: coefficient.estimates}}).

\subsection{\it Expected Values of Anomalous Couplings}

The various couplings in Equations~\ref{wwv} and \ref{zgv} can be 
explicitly calculated within
the SM or any of its extensions. In this subsection we briefly review
the results for various models. 

Most of the existing calculations refer
to the $WWV$ vertices and concentrate on the
$ \Delta \kappa_V , \; \tilde \kappa_V $, and $ \lambda_V$ couplings. 
Though the specific values obtained are model dependent,
they are all in compliance with the
estimates given in Section~{\ref{sec: coefficient.estimates}}.

A careful calculation of the couplings in Equation~\ref{wwv} must preserve
the gauge invariance of the model. The most careful computations of which
we are aware~\cite{pinch} carefully preserve
gauge invariance and are devoid of pathologies such as infrared
divergences.
The results of the calculations are presented in Table~\ref{tab:table2}.

Note that the extremely small SM value for $ \tilde \kappa_\gamma $
is due to the fact that
the electric dipole moment of the $W$ boson in the SM vanishes at both the
one- and two-loop levels~\cite{sm13,sm20}. This parameter takes the 
larger values $8 \times 10^{-10}$ in a model with mirror 
fermions~\cite{sm13} and $5 \times 10^{-3}$ in models with a 
fourth generation~\cite{various10}.
The parameters in Equation~\ref{zgv} have received much less attention. The only
calculations of which we are aware predict values of $ h_3 \sim 10^{-6}$ 
for a two
Higgs doublet model~\cite{chang.et.al.}, and  $ h_3^Z \sim 1.3
\times 10^{-6} $ for the SM top quark loop contribution~\cite{barroso}.

\section{\rm \large ASSOCIATED GAUGE BOSON PRODUCTION\\
AT \ppbar\ COLLIDERS}
\label{sec:pheno}

At leading order, associated production of gauge bosons takes place
via the Feynman diagrams shown in Figure~\ref{fig:feynman1}. The \wg\
process has the highest cross section among the gauge boson pair
production processes at the Tevatron.  Many authors have discussed the
use of \wg\ production at hadron colliders to probe anomalous \wwg\
couplings~\cite{wg-lit}. A tree level calculation of the
\wg\ cross section with anomalous couplings parametrized in the most
general model-independent way, using the effective Lagrangian approach,
has been performed by Baur and co-workers~\cite{Baur&Zeppenfeld-wgmc,
Baur&Berger-wgmc}.

In Figure~\ref{fig:feynman1}\textit{a} and \ref{fig:feynman1}\textit{b} the $t$- and $u$-channel Feynman
diagrams for $p \bar p \to \ell \nu \gamma$ correspond to photon
bremsstrahlung from an initial-state quark. The \wwg\ coupling appears
in the $s$-channel process, Figure~\ref{fig:feynman1}\textit{c}.  Events in which a
photon is radiated from the final state lepton from single \W\ boson decay
(Figure~\ref{fig:feynman2}) also result in the same $\ell \nu \gamma$
final state.

The contributions from anomalous couplings to the helicity amplitudes
for $p \bar p \to W \gamma$ can be written as~\cite{Baur&Zeppenfeld-wgmc}
\begin{eqnarray}
\Delta {\cal M}_{\pm 0} = {e^2 \over {\rm sin} \, \theta_W}
{\sqrt{\hat s} \over 2 m_W} 
\left[
\Delta\kappa_\gamma + \lambda_\gamma \mp i(\tilde\kappa_\gamma
+\tilde\lambda_\gamma)
\right]
{1 \over 2} (1 \mp {\rm cos} \, \Theta)  \nonumber \\
\Delta {\cal M}_{\pm \pm} = {e^2 \over {\rm sin} \, \theta_W}
{1 \over 2} 
\left[
{\hat s \over m^2_W}
(\lambda_\gamma \mp i\tilde\lambda_\gamma)
+ (\Delta\kappa_\gamma \mp i\tilde\kappa_\gamma) 
\right]
{1 \over \sqrt{2}} {\rm sin} \, \Theta,
\label{eqn:wg-hel}
\end{eqnarray}
where the subscripts of $\Delta {\cal M}$ denote the photon and $W$
helicities (the quark helicities are fixed by the $V-A$ structure of
the $Wq \bar q$ coupling), and $\Theta$ denotes the scattering angle of
the photon with respect to the quark direction, measured in the \wg\
rest frame. From these expressions we see several important features:

\noindent
(\textit{a}) the cross section increases quadratically with the anomalous
coupling parameters;

\noindent
(\textit{b}) due to the $\sqrt{\hat s} / m_W = \hat \gamma_W$ factors 
in these expressions, the effects of anomalies in the \wwg\ vertex are
enhanced at large parton subprocess energies. Therefore, a typical
signature for anomalous couplings is a broad increase in the 
\wg\ invariant mass at large values of $\hat s = m_{W\gamma}$;

\noindent
(\textit{c}) the sensitivity to $\lambda_\gamma$ will be higher than for $\Delta
\kappa_\gamma$, because of the factor ${\hat \gamma_W}^2$ multiplying 
$\lambda_\gamma$ in Equation~\ref{eqn:wg-hel}.

A striking feature of the $p \bar p \to W \gamma$ process is the
prediction of radiation zeros in all the helicity amplitudes for SM
\wwg\ couplings. For $u \bar d \to W^+ \gamma$ the amplitudes vanish at ${\rm
cos} \, \Theta = - 1/3$.
In the presence of anomalous couplings the
radiation zero is partially eliminated. This is evident from
Equation~\ref{eqn:wg-hel} since all amplitudes are finite for 
nonzero anomalous couplings and ${\rm cos}\, \Theta = - 1/3$. 
Consequently the average photon \pt\ increases
considerably in the presence of anomalous couplings and therefore the photon
\pt\ distribution is particularly sensitive to anomalous
couplings. This effect, illustrated in Figure~\ref{fig:ptgamma}\textit{a},
can be understood because anomalous couplings
contribute through the $s$-channel diagram so their effects are
evident predominantly in the central (low-rapidity) region.
The photon \pt\ is the quantity used in the \d0\ and CDF experiments to 
search for anomalous couplings because it is easier to measure
than the \wg\ invariant mass. The latter requires a knowledge of the
neutrino longitudinal momentum, which cannot be measured at a \ppbar\ collider.

The Feynman diagrams for $p \bar p \to Z \gamma$ are shown in
Figure~\ref{fig:feynman1}. Since the $Z$ has zero electric charge and
zero weak isospin, the \zzg\ and \zgg\ couplings are zero at tree level
in the SM, and the $s$-channel diagram only contributes in the
presence of anomalous couplings. As a result there
is no radiation zero in \zg\ production.  Moreover, in the SM the
ratio of the \zg\ cross section to the \wg\ cross section rises with
increasing minimum photon \pt\ due to suppression of \wg\ by the
radiation zero, unlike the ratio for $Zj/Wj$, which is approximately
independent of the minimum jet \pt~\cite{Baur&Errede&Ohnemus}.

The leading-order helicity amplitudes and cross section for \zg\
production have been evaluated~\cite{Baur&Berger-zgmc}. For
anomalous couplings the $s$-channel diagram contributes, resulting in
events with higher average photon \pt, as shown in
Figure~\ref{fig:ptgamma}\textit{b}. Defining $\hat \gamma_Z = \sqrt{\hat
s}/m_Z$, it is found that the terms in the anomalous contributions to
the helicity amplitudes are multiplied by factors of $\hat \gamma_Z^3$
for $h^V_1$ and $h^V_3$ and $\hat \gamma_Z^4$ for $h^V_2$ and $h^V_4$.
Thus the growth with $\hat s$ is faster than for \wg\ production and
the experimental limits are more sensitive to the choice of form
factor scale $\Lambda_{FF}$. Finally, we note that \zg\ production has been
studied experimentally using the final states $\ell^+ \ell^- \gamma$ and
$\nu \bar \nu \gamma$. The latter has a higher branching fraction, and
since there is no charged lepton involved, the final-state radiation
process (as in Figure~\ref{fig:feynman2}) is absent. 

The \ww\ and \wz\ production processes are also sensitive to anomalous
couplings~\cite{wwxs_lo}. The \wwg\ and \wwz\ anomalous couplings
enter via the $s$-channel diagram, Figure~\ref{fig:feynman1}\textit{c}. The
effects are an increase in the average value of the invariant mass of
the boson pair ($m_{WW}$ and $m_{WZ}$) and of the boson transverse
momentum ($p_T^W$ and $p_T^Z$). 

The \ww\ production process is sensitive to both the \wwg\ and
\wwz\ couplings. When deriving limits on the anomalous couplings it is
therefore customary to make assumptions about the relations between
the \wwg\ and \wwz\ coupling parameters in order to reduce the number
of free parameters. For example, one can assume
SM \wwg\ couplings and derive limits on the \wwz\ coupling parameters
or vice versa. The sensitivity to the \wwz\ couplings is higher due to
the overall coupling $g_{WWZ}$ for the \wwz\ vertex, which is larger
than the corresponding factor $g_{WW\gamma}$ for the \wwg\ vertex.
Alternatively, one can assume equal \wwz\ and \wwg\ couplings
($\Delta\kappa_\gamma = \Delta\kappa_Z$, etc), or use the HISZ
assumption (see Section {\ref{sec:lagrangians}}).  The \wz\ production process
has the advantage of being sensitive only to the \wwz\ coupling, but
it has a smaller cross section. The SM cross sections are $\sigma_{WW}
= 9.5$~pb~\cite{wwxs_nlo} and $\sigma_{WZ} = 2.5$~pb~\cite{wzxs_nlo}
at next-to-leading order.

An important consideration for the experimental study of diboson
production at the Tevatron is the effects of higher order
c on the cross sections and kinematic
distributions. Next-to-leading--order calculations have been performed
for SM and anomalous couplings for \wg~\cite{wgzgxs_nlo, wg-nlo},
\zg~\cite{zg-nlo}, \ww~\cite{wwxs_nlo, ww-nlo}, and
\wz~\cite{wzxs_nlo, wz-nlo} production at hadron colliders. At the Tevatron
energy, the next-to-leading--order cross sections are generally a factor of $\sim 30$\%
higher than the leading-order calculations. The shapes of the kinematic
distributions are, to a good approximation, unchanged compared 
with leading order. 
Therefore, the analyses described below
have used leading-order Monte Carlo event generators with a
K-factor of $1 + \frac{8}{9}\pi \alpha_s(M_W^2) \approx 1.34$
to approximate the effects of the QCD corrections. The transverse
momentum of the diboson system is modeled based on the
observed $W$ $p_T$ spectrum in inclusive $W \to e \nu$ events.

Experimental limits on the anomalous couplings are derived using
the form factor ansatz described in Section
{\ref{sec: form.factors}}. The
motivation for the choice of the form factor scale $\Lambda_{FF}$ is
illustrated in Figure~\ref{fig:ff-scale}\textit{a}, which shows an old
experimental limit and the corresponding unitarity
limit~\cite{baur-unitarity}
as a function of
$\Lambda_{FF}$ for the
\zzg\ coupling parameter $h^Z_{30}$. At large $\Lambda_{FF}$ the unitarity
limit becomes more stringent than the experimental limit. Therefore,
$\Lambda_{FF}$ is chosen to be as large as possible consistent with 
unitarity as indicated by the vertical arrow in
Figure~\ref{fig:ff-scale}\textit{a}. In practice, round
numbers (e.g. $\Lambda_{FF} = 500$~GeV) are used
to allow easy comparison of results between different experiments.
For \wg\ production, the limits depend only weakly on the form factor
scale for $\Lambda_{FF}$ above about 500~GeV, as shown in 
Figure~\ref{fig:ff-scale}\textit{b}.

\section{\rm \large THE TEVATRON COLLIDER AND DETECTORS}
\subsection{\it The Tevatron Proton--Antiproton Collider}

A schematic of the Fermilab accelerator complex~\cite{accelerators} is
shown in Figure~\ref{fig:Tevatron}. In the Linac, 18~keV $H^-$ ions from
a Cockroft-Walton electrostatic generator are accelerated to an energy
of 200~MeV. The electrons are then stripped off and the remaining
protons are injected into the Booster, where they are accelerated to
8~GeV. They are then transferred to the Main Ring, a 1~km--radius
synchrotron located in the same tunnel as the Tevatron.
Protons are accelerated to 150~GeV in the Main Ring and injected into
the Tevatron.  The Tevatron~\cite{Tevatron} was the first large
accelerator to use superconducting magnets for the main guide field.
It accelerates protons and antiprotons to a final energy of 900~GeV.

The Main Ring also provides a beam of 120~GeV protons, which are
extracted and strike a target, producing antiprotons with a peak energy
of 8~GeV \cite{pbar-source}. The antiprotons are stochastically cooled
and stacked in the Debuncher and Accumulator and transferred to the
Main Ring for injection into the Tevatron.

One of the main limitations on the achievable luminosity is the number
of antiprotons in the accelerator. During Run~I (1992--1996), the
Tevatron was operated with six antiproton (and six proton) bunches and
with $\approx$$7 \times 10^{10}$ antiprotons per bunch. This led to
peak luminosities of $\approx$$3 \times 10^{31}$~cm$^{-2}$~s$^{-1}$.
The time interval between bunches, which determined the interval
between collisions in each detector, was 3.5~$\mu$s during Run~I.

\subsection{\it The CDF and \d0\ Detectors}

The CDF detector \cite{CDF-detector} is shown in Figure~\ref{fig:CDF-detector}. The tracking system consists of an inner
silicon microstrip vertex detector, a set of time projection chambers,
and an outer central tracking chamber, covering $|\eta| < 1.1$. 
The pseudorapidity $\eta$ is defined as $\eta = -{\rm ln}\left[{\rm
tan} \, \theta/2 \right]$, where $\theta$ is the polar angle with
respect to the beam axis.

These
detectors provide a measurement of the transverse momentum \pt\ of charged
particles with a resolution of $\sigma (p_T) / p_T \approx$ $\sqrt{(0.9
p_T)^2 + (6.6)^2} \times 10^{-3}$, where $p_T$ is in GeV/$c$.  The
central electromagnetic and hadronic calorimeters consist of
lead-scintillator and steel-scintillator sampling detectors,
respectively.  The energy resolution for $|\eta| < 1.1$ is
$\sigma(E)/E$ $\approx$14\%$/\sqrt{E}$ for electrons and
$\approx$(50 to 75)\%$/\sqrt{E}$ for isolated pions
where $E$ is in GeV. In the forward
region ($1.1 < |\eta| < 4.2$) the calorimeters use proportional
chambers and have energy resolution $\approx$25\%$/\sqrt{E}$ for
electrons and $\approx$110\%$/\sqrt{E}$ for isolated pions.  The
calorimeter projective tower segmentation in the central region is
$0.1 \times 0.26$ in $\eta \times \phi$, where $\phi$ is the
azimuthal angle, while in the forward
region it is $0.1 \times 0.09$.  The central muon system consists
of a set of drift chambers and steel absorbers covering the region
$|\eta| < 1.0$.

The \D0\ detector~\cite{D0-detector} consists of three main systems
(Figure~\ref{fig:D0-detector}).  The central drift chamber and forward
drift chambers are used to identify charged tracks for $|\eta| < 3.2$
and to measure the position of interaction vertices along the
direction of the beam.  The calorimeter consists of 
uranium/liquid-argon sampling detectors with fine 
segmentation, and is divided into a
central and two endcap cryostats covering $|\eta| < 4.4$. The energy
resolution of the calorimeter is $\approx$15\%$/\sqrt{E}$ for
electrons and $\approx$50\%$/\sqrt{E}$ for isolated pions. 
The calorimeter towers subtend $0.1 \times 0.1$ in
$\eta \times \phi$, segmented longitudinally into four electromagnetic
(EM)
layers and four or five hadronic layers. In the third EM layer, at the
EM shower maximum, the cells are $0.05 \times 0.05$ in $\eta \times
\phi$.  The muon system consists of magnetized iron toroids with one
inner and two outer layers of drift tubes, providing coverage for
$|\eta| < 3.3$.

\section{\rm \large DETECTION OF ASSOCIATED GAUGE BOSON\\
PRODUCTION}

In the Tevatron Run~I analyses of diboson events, the following final states are 
considered:\\

\noindent
\tab 1. $W \gamma \to \ell \nu \gamma$\\
\tab 2. $WW \to \ell \nu~\ell \nu$\\
\tab 3. $WW/WZ \to \ell \nu~q \bar q$\\
\tab 4. $WZ \to q \bar q~\ell^+ \ell^-$\\
\tab 5. $Z \gamma \to \ell^+ \ell^- \gamma$\\
\tab 6. $Z \gamma \to \nu \bar \nu \gamma$\\

\noindent
where $\ell = e$~or~$\mu$. Except for $Z \gamma \to \nu \bar \nu \gamma$,
only electron and muon decays of the $W$ and $Z$ have been studied, as
they provide a unique experimental signature of high-$p_T$ isolated
leptons. Although CDF have reported a $WZ \to e \nu~ee$ candidate
event and a $ZZ \to \mu \mu~\mu \mu$ candidate event in their 
data~\cite{cdf_wwlvjj1b},
these modes have not been studied so far due to the much lower cross
section times branching fraction (less than one event is expected in
each mode in Run~I).
Detection of charged leptons, neutrinos, photons, and jets
from the six processes listed above is discussed in the following sections.

\subsection{\it Detection of Leptonic W and Z Decays}

Electrons from $W$ or $Z$ decays are identified as tracks in the
tracking chambers pointing to the centroid of a shower in the
electromagnetic calorimeters. To discriminate against charged hadrons,
the profile of energy deposition in the calorimeter and the fraction
of electromagnetic energy to hadronic energy must be consistent with
electron test beam studies and with clean samples of electrons
obtained from collider data.
The electron calorimeter showers are
required to be isolated from nearby energy deposition in the
calorimeter. CDF also use track isolation.

Muons are reconstructed as tracks in the muon chambers.  Additional
identification requirements are used to reject cosmic ray muons and
hadrons, which interact in the calorimeters.  The impact parameter of
the muon track from the beamline and from the interaction vertex
$z$-position must be consistent with that of a particle originating
from the hard collision, and the energy deposited in the calorimeters
must be characteristic of a minimum ionizing particle. In \d0\ the
muon must also be coincident with the beam crossing time.  In CDF the
muon transverse momentum is measured using the central tracking
chamber; in \d0\ it is measured using the muon chambers and toroidal
field with a resolution of $\sigma(1/p) \approx 0.18(p-2)/p^2 \oplus
0.008$, with $p$ in GeV/$c$. Therefore, the CDF measurement of
momentum is more precise.  The muon tracks are required to
be isolated from nearby jets and from energy deposition in the
calorimeters. CDF also uses track isolation in the central tracking chambers.

Neutrinos from $W$ decays are inferred from the missing transverse
energy in an event. The neutrino $p_T$ is calculated from the missing energy
in the calorimeters and the transverse momentum of muons in the event (if any).

\subsection{\it Photon Detection}
\label{sec:photon-detection}

Photons must satisfy the same selection criteria as electrons,
except that the electromagnetic shower must not be accompanied by a
matching track. In some of the \d0\ analyses, photon candidates
containing hits in the region of the central tracking chambers
between the interaction vertex and the EM cluster centroid
are rejected too, as this indicates an unreconstructed track.

In the \wg\ and \zg\ analyses, an important source of background
originates from $W$~+~jet and $Z$~+~jet events, where jet fragmentation
fluctuations lead to a single neutral meson such as a $\pi^0$ carrying
most of the energy of the jet. For meson transverse energies above
about 10~GeV, the showers from the two decay photons coalesce and mimic a
single photon shower in the calorimeter.

To estimate these backgrounds it is necessary to calculate the
probability that a jet ``fakes'' a photon, $P(j \to
``\gamma$''). In both \d0\ and CDF this is done using a sample of
multijet events obtained from jet triggers, which are independent of
the triggers used to select the \wg\ and \zg\ signal events. The
probability is determined as a function of the $E_T$ of the jet by
measuring the fraction of nonleading jets in the multijet sample that
passes the photon identification requirements.
To avoid trigger biases associated with the calorimeter energy
response at trigger threshold, only the nonleading jets are used,
i.e. those jets that did not fire the trigger.

The multijet samples also contained genuine direct photons,
predominantly from gluon Compton scattering. In \d0\ the fraction of
such direct photons in the sample is determined using the energy
deposited in the first layer of the EM calorimeter. Since meson decays
produce two photons, which can independently convert to an $e^+e^-$
pair in the calorimeter, showers originating from mesons start earlier
than single photon showers and produce more energy in the first layer
of the calorimeter.  In CDF the transverse shape of the shower at the
shower maximum is used, since on average it is broader for meson decay
showers than for single photon showers.  The resulting
probability $P(j \to ``\gamma$'') is found to be in the range
$\approx 10^{-4} - 10^{-3}$, depending on the photon identification
requirements and on $E_T$.

\subsection{\it Jet Detection}
\label{sec:jet-detection}

In the processes $WW/WZ \to \ell \nu~q \bar q$ and $WZ \to q \bar q~\ell^+ \ell^-$, either a $W$ or a $Z$ decays to a $q
\bar q$ pair, which hadronizes to form jets.  Production of single $W$
or $Z$ bosons with a subsequent decay to two jets has not been
observed at the Tevatron due to the overwhelming background
from two-jet events. In the region of the $W$ and
$Z$ masses, this background consists mainly of gluon jets, which are
indistinguishable from quark jets on an event-by-event basis.

Similarly, \ww\ and \wz\ production in which one boson decays to
leptons and the other to a $q \bar q$ pair has not been isolated from
the very large background due to $W \, + \, jj$ or $Z \, + \, jj$ 
production and
multijet production. However, to retain good acceptance for the
SM signal and anomalous \ww\ and \wz\ production while
minimizing the backgrounds, the analyses require events to contain two
jets that form an invariant mass consistent with the $W$ or $Z$ mass.
The jet energy resolutions are typically
$\sigma(E)/E\approx$100\%$/\sqrt{E}$ with $E$ in GeV.  The dijet
invariant mass resolution is approximately 10~GeV$/c^2$ for $m_{jj} =
80$~GeV$/c^2$.

Jets are detected in the CDF and \d0\ hadron calorimeters using a
fixed cone clustering algorithm with cone radius $R = \sqrt{\Delta
\eta^2 + \Delta \phi^2}$. For the analyses under consideration here,
cone sizes of $R=0.3, 0.4$, and 0.5 have been used. For smaller cone
sizes, fragmentation effects cause particles to be lost outside the
clustering cone, resulting in poorer energy resolution.  For larger
cone sizes, more energy associated with the underlying event is
included within the cone, also resulting in poorer energy resolution;
furthermore, jets close together in $\eta - \phi$ space tend to be merged into
one jet. The latter effect results in low efficiency for detecting the
$W \to q \bar q$ and $Z \to q \bar q$ decays, especially at high boson
transverse momentum, since the opening angle of the two jets decreases
as the $p_T$ of the boson rises. Typically, the efficiency starts to
drop off for $p_T(jj) >$~200 to 300~GeV$/c$.

%

\section{\rm \large ANALYSIS AND RESULTS}
\label{sec:analysis}

\subsection{\it \wg\ Analysis Results}
\label{sec:wg_anal}

The published results on \wg\ production at the Tevatron are
from the CDF analysis of data from Run~Ia \cite{cdf_wg1a_prl,
cdf_wgzg1a_prd} (1992--1993) and the \d0\ analyses of data from Runs Ia and
Ib~\cite{d0_wg1a_prl, d0_diboson_1a_prd, d0_wg1b_prl} (1994--1995). Also
described in this section are the preliminary results from the CDF analysis of
data from Run Ib~\cite{cdf_wg1b}.

In the analyses, a high $p_T$ electron or muon is required (see
Table~\ref{table:wg_evtsel}), accompanied by large missing transverse
energy, indicating the presence of a \w\ boson.  A high \pt\ isolated
photon is also required, with $p_T > 7$~GeV$/c$ for CDF and $p_T >
10$~GeV$/c$ for \d0. The higher $p_T$ cut used by \d0\ results in a
lower acceptance, but does not reduce the sensitivity to anomalous
couplings, because anomalous couplings would result in events
with higher $p_T$ photons compared with the SM.  The
photon is required to be separated from the lepton by $\Delta R_{\ell
\gamma} > 0.7$ units in $\eta - \phi$ space, which reduces the background
from radiative \w\ decays.  Photons and electrons are detected in
the pseudorapidity range $|\eta| < 1.1$ for CDF and $|\eta| < 1.1$ or
$1.5 < |\eta| < 2.5$ for \d0.  This results in a higher geometrical
acceptance for \d0.

The backgrounds
are
from the following sources: (\textit{a}) $W$~+~jet production, where the jet fluctuates to a neutral
meson such as a $\pi^0$ which decays to two photons; (\textit{b}) \zg\ events
in which one of the leptons from the \z\ decay is not reconstructed;
(\textit{c}) $W(\tau \nu)\gamma$ production with the decay $\tau \to \ell \nu
\nu$; and (\textit{d}) processes (labeled $\ell e X$) that produce missing
transverse energy, a high-$p_T$ lepton, and an electron with an
unreconstructed track.  The dominant background is from (\textit{a}). This
background
is estimated from the observed $E_T$ spectrum of jets in the
inclusive $W \to \ell \nu$ data samples and from the measured probability
for a jet to fake a photon (see {Section~{\ref{sec:photon-detection}}}).
The smaller backgrounds from (\textit{b}) and
(\textit{c}) are estimated using Monte Carlo simulations.  Background (\textit{d})
is significant only in the \d0\ analysis due to the small but nonzero
inefficiency for reconstructing tracks associated with electrons.  The
sources of this background are from $t \bar t$ and $WW$ pair
production with a subsequent $W \to \ell \nu$ decay, and in the
electron channel, high-$p_T$ $Z \to ee$ and multijet production.

Theoretical predictions of \wg\ production are made based on the
leading-order Monte Carlo program of Baur \&
Zeppenfeld~\cite{Baur&Zeppenfeld-wgmc, Baur&Berger-wgmc}
(see {Section~\ref{sec:pheno}}). The
efficiencies and acceptances of the CDF and \d0\ detectors are
modeled using fast Monte Carlo programs that include geometrical
acceptances and  smearing effects due to detector resolutions.

Table~\ref{table:wg_nevts} compares the numbers of signal events after background subtraction
with the SM predictions. The number
of events is of the order of 100 for each experiment.  The \d0\ measured cross section times branching fraction (with $E_T^\gamma >
10~\GeV$ and $\Delta R_{\ell \gamma} > 0.7)$ is $\sigma (W \gamma) \times
B(W~\to~\ell \nu) = 11.3^{+1.7}_{-1.5}~{\rm (stat)} \pm 1.5$~(syst)~pb
compared with the SM prediction of $\sigma (W \gamma) \times B(W \to
\ell \nu) = 12.5 \pm 1.0$~pb.  Figure~\ref{fig:wghists_d0} shows the \d0\
$p_T^\gamma$ distribution for the observed candidate events together
with the SM signal prediction plus the sum of the estimated
backgrounds. The number of observed events and the shapes of the
distributions show no deviations from the expectations.

In both experiments, limits on the \wwg\ vertex coupling parameters
are obtained from a binned maximum likelihood fit to the photon \pt\
distribution. The likelihood function is given by
\begin{eqnarray}
P(\sigma | n)&=& \int_{0}^{\infty} d{s_i}
		 \int_{0}^{\infty} d{b_i} 
                 \prod_i \frac{e^{-(b_i + s_i)}
                 (b_i + s_i)^{n_i}}{n_i!} \nonumber \\
	     && \times G(s_i;s_{i0},\sigma_{s_i})
                       G(b_i;b_{i0},\sigma_{b_i}),
\end{eqnarray}
\noindent
where $b_i + s_i$ is the predicted number of events in
the $i^{\rm th}$ bin, $b_i$ is the estimated background in the $i^{\rm
th}$ bin, $s_i = {\cal L}\epsilon_i\sigma_i$ is the predicted number
of signal events in the $i^{\rm th}$ bin,
${\cal L}$ is the integrated luminosity, $\epsilon_i$ is the
efficiency for the $i^{\rm th}$ bin, $\sigma_i$ is the \wg\ cross
section prediction for the $i^{\rm th}$ bin, and $n_i$ is the observed
number of events in the $i^{\rm th}$ bin.  In the above expression,
the Poisson probability for each bin is convolved with two Gaussian
distributions $G$, which represent the uncertainties in the 
background estimate and the predicted number of signal events.
This method \cite{cousins} incorporates these uncertainties into the confidence
interval calculation using a Bayesian statistical approach, while the
Poisson probability is treated classically.
The quantity $s_i = s_i(\Delta \kappa_\gamma, \lambda_\gamma)$ depends on the
anomalous coupling parameters $\Delta \kappa$ and $\lambda$. 
To exploit the fact that anomalous coupling contributions lead to an
excess of events at high photon $p_T$, a high $p_T$ bin in which no
events were observed is explicitly included in the histogram. The
nonobservation of events in this bin carries information on the
anomalous couplings~\cite{landsberg-thesis}.

For the \wg\ analysis and the analyses described in subsequent sections,
the uncertainties are typically 
$\approx$10\% from the errors in the measured
detection efficiencies, $\approx$5\%
from the choice of parton distribution function (pdf),  $\approx$1\%
from varying the $Q^2$ at which the pdf's are evaluated,  $\approx$5\%
from the modeling of the diboson transverse momentum, and 
$\approx$6\% from the integrated luminosity measurement error.
The uncertainty in the background estimates varies from 12\% to 30\%
depending on the particular analysis channel.

In the electron channel Run~Ib analysis, \d0\ requires the
electron-photon-neutrino transverse cluster mass to
be~$>90$~GeV$/c^2$. This requirement suppresses radiative \w\ decays and
increases the sensitivity to anomalous couplings by about 10\%.
Figure~\ref{fig:wglimits} shows the 95\% confidence level (CL) limits
in the $\Delta \kappa_\gamma - \lambda_\gamma$ plane, for a form 
factor scale of
$\Lambda_{FF} = 1.5~\TeV$. Varying only one coupling at a time
from its SM value, the
following limits are obtained at the 95\% CL\footnote{
When only one parameter is varied, the 95\% CL limits are obtained
from the points where the log-likelihood function has fallen by
1.92 from its  maximum. For two free parameters, e.g. in a
plot of $\lambda_\gamma$ vs $\Delta\kappa_\gamma$, 
the 95\% CL limits are obtained
from the points where the log-likelihood function has fallen by
3.00 from its  maximum. These are sometimes referred to as 1-d and
2-d limits.}:
\begin{tabbing}
{\d0}:~~~~~~~~~~~\=$-0.93 < \Delta \kappa_\gamma < 0.94$\\ 
     \>$-0.31 < \lambda_\gamma < 0.29$\\
CDF: \>$-1.8 < \Delta \kappa_\gamma  < 2.0$\\
     \>$-0.70 < \lambda_\gamma < 0.60$
\end{tabbing}
The possibility of a minimal U(1)$_{\rm EM}$-only coupling ($\kappa =
\lambda = 0$) indicated by the solid circle in Figure~\ref{fig:wglimits} is
ruled out at the 88\% CL by the \d0\ measurement.
Figure~\ref{fig:qmulimits} shows the limits on a plot of
$W$ boson electric quadrupole moment $Q_W$ vs magnetic 
dipole moment $\mu_W$.

The ${\rm cos} \, \Theta$ distribution in the data is consistent with
the SM prediction \cite{cdf_wg1b}, where $\Theta$ is the scattering
angle of the photon with respect to the quark direction, measured in
the \wg\ rest frame.  However, at present the integrated luminosity is
too low to establish the presence of the radiation zero.

\subsection{\it $WW \to \ell \nu \ell^\prime \nu^\prime$ Analysis Results}

The CDF and \d0\ experiments have searched for $W^+W^-$ production in the
dilepton decay modes $e \nu e \nu$, $e \nu \mu \nu$, and $\mu \nu \mu
\nu$ \cite{d0_1a_wwdilep, cdf_wwdilep, d0_1b_wwdilep}. The event
selection requirements are summarized in
Table~\ref{table:wwdilep_evtsel}.  The CDF and \d0\ analyses each
require two isolated leptons plus missing transverse energy, using
similar selection criteria. 

Background from top quark pair production ($p \bar p \to t \bar t + X
\to W^+ W^- b \bar b + X$) is suppressed by removing events that contain
hadronic energy in the calorimeters.  \d0\ requires the vector sum of
the $E_T$ from hadrons ${\vec E}_T^{\rm had}$, defined as ${\vec
E}_T^{\rm had} = -({\vec E}_T^{\ell 1} + {\vec E}_T^{\ell 2}
~+~$\etmisv$)$, to be less than 40~GeV. For $W^+W^-$ events, gluon radiation
and detector resolution give rise to small values of ${\vec E}_T^{\rm
had}$ compared with $t \bar t$ events, where the main contribution is
from the $b$-quark jets from the $t$-quark decays.
This cut reduces the background from $t \bar t$ production by
a factor of more than four for $m_t = 170$~GeV$/c^2$, with an
efficiency of 95\% for SM $W^+W^-$ events. CDF suppresses 
the $t \bar t$ background by removing events containing
any jet with $E_T > 10$~GeV.

To discriminate against backgrounds from $Z \to \tau^+ \tau^-$ and the
Drell-Yan processes $\gamma / Z \to e^+e^-, \mu^+\mu^-$, a \etmis\ cut is
applied (Table~\ref{table:wwdilep_evtsel}). Events are also rejected
if the \etmis\ vector points along the direction of a lepton or
opposite to the direction of a lepton (within $20^\circ$) and the
\etmis\ is less than 50~GeV. Finally, events with a
dilepton mass within the limits 75~GeV$/c^2 < m_{\ell^+\ell^-} <
105$~GeV$/c^2$ are rejected.

In the \d0\ analysis based on an integrated luminosity of
97~pb$^{-1}$, five events pass the event selection criteria and the
total estimated background is $3.1 \pm 0.4$ events.
This leads to an upper limit on the cross section for $p \bar
p \to W^+W^-$ of 37.1~pb at the 95\% CL.

In the CDF analysis based on 108~pb$^{-1}$ of data,
the event selection also yields five events, but with an estimated
background of only $1.2 \pm 0.3$ events. The probability that the observed
events correspond to a fluctuation of the background is 1.1\%.
The $W^+W^-$ cross section is measured to be $\sigma (p \bar p \to W^+W^-) =
10.2^{+6.3}_{-5.1} \pm 1.6$~pb. 
This is in good agreement with the 
next-to-leading--order cross section for SM \w\ pair production
calculated by Ohnemus~\cite{wwxs_nlo}, which gives 
the result $\sigma_{SM} (p \bar p \to W^+W^-) = 9.5$~pb.
Because of the higher signal-to-background ratio, CDF is able to make a
measurement of the cross section rather than set an upper limit as \d0 does.

The \w\ pair production process is sensitive to both the \wwg\ and \wwz\
couplings, since the $s$-channel propagator can be a $\gamma$ or a 
\z. Anomalous couplings result in a higher cross section and an
enhancement of events with high \pt\ \w\ bosons. In the CDF analysis
the total cross section is used in setting limits, while in the \d0\
analysis a binned maximum likelihood fit is performed to the measured
\pt\ spectra of the two leptons in each event~\cite{d0_1b_wwdilep}.
This technique is
similar to that described for the \wg\ analysis previously. However, there
is a correlation between the \pt\ of one lepton and the \pt\ of the
other lepton in the same event because the two $W$ bosons are boosted
by approximately the same amount in opposite directions (${\vec p}_T^{W^+}
\approx -{\vec p}_T^{W^-}$). This correlation is stronger for larger anomalous
couplings because of the higher \pt\ of the $W$ bosons in these
events. To account for this correlation, two-dimensional bins in the \pt\ 
of one lepton vs the \pt\ of the other lepton are used.  Use of this
kinematic information provides significantly tighter constraints on
anomalous couplings than those obtained from the measurement of the cross
section alone. Both experiments use a tree-level Monte Carlo
program~\cite{wwxs_lo} to generate $W^+W^- \to \ell^+ \nu
\ell^- \bar \nu$ events as a function of the coupling parameters. 

Varting only one coupling at a time and
assuming $\Delta \kappa_Z = \Delta \kappa_\gamma$ and $\lambda_Z =
\lambda_\gamma$, the \d0\ results based on 97~pb$^{-1}$ of data using
the kinematic likelihood fit method yield the following 95\% CL
limits for a form factor scale of $\Lambda_{FF} = 1.5$~TeV:
\begin{center}
$-0.62 < \Delta \kappa < 0.77$ \\
$-0.52 < \lambda < 0.56$.
\end{center}
\noindent
%
%
The limits obtained by CDF using the cross section alone are slightly
looser than the \d0\ limits and are given in reference~\cite{cdf_wwdilep}.
%

\subsection{\it $WW/WZ \to \ell \nu jj$ and $WZ \to jj \ell \ell$
Analysis Results}

In the lepton-plus-jets analyses CDF~\cite{cdf_wwlvjj1a_prl,cdf_wwlvjj1b}
and \d0~\cite{d0_wwlvjj1a_prl,d0_wwlvjj1b_prl}
search for candidate
$WW/WZ \to \ell \nu jj$ events containing a high \pt\ lepton, missing
\et\, and two jets with invariant mass consistent with the \w\ or \z\
mass (taking into account the dijet mass resolution of $\approx
10$~GeV$/c^2$). The event selection requirements are given in
Table~\ref{table:wwlvjj_evtsel}.  CDF also accepts events with two
charged leptons and two jets resulting from $p \bar p \to WZ \to
jj \ell^+ \ell^-$. The event selection is similar in all respects,
except that a second lepton is required in place of the missing \et\
requirement.
CDF has analyzed the electron and muon decay channels, while \d0\
has so far only analyzed the $e \nu jj$ final state.

The CDF and \d0\ analyses follow similar lines.  After applying
the lepton, missing \et\, and jet requirements, the invariant mass of
the dijet system is histogrammed. For events containing more than two jets,
CDF takes the two leading jets whereas \d0\ uses the dijet combination with
the largest invariant mass.
Figure~\ref{fig:wwlvjj_cdfhists}\textit{a} shows the resulting histogram for
the CDF Run~Ib preliminary analysis. The dijet mass cut (see 
Table~\ref{table:wwlvjj_evtsel}), which selects the events falling within the shaded band in the
figure, is then applied.  The transverse momentum of the two-jet system for this subset
of events is shown in Figure~\ref{fig:wwlvjj_cdfhists}\textit{b}.

Jet cone radii of $R = 0.3, 0.4$, and 0.5 were used in the analyses
(see
Table~\ref{table:wwlvjj_evtsel}).
We discuss the motivations for these choices in 
{Section~{\ref{sec:jet-detection}}

As shown in Figure~\ref{fig:wwlvjj_cdfhists}, the data are dominated by
background, mainly from $W~+ \geq 2$~jets events with $W \to e \nu$
and (in \d0) multijet production where one jet is misidentified as an
electron and there is significant (mismeasured) missing \et.
However, at large values of $p_T^W$ the backgrounds are relatively
small and it is  predominantly in this region where anomalous couplings
enhance the cross section. This is the key to obtaining limits on the
anomalous couplings. The main difference between the analyses is that
CDF applies a cut on the boson transverse momentum ($p_T^W = p_T^{jj} >
200$~GeV$/c$) and extracts limits on the anomalous couplings from the
number of events surviving the cut, whereas \d0\ uses a binned likelihood
fit to the $W$ boson \pt\ spectrum. The latter technique is analogous
to the fit to the $p_T^\gamma$ distribution in the \wg\ analysis.

The CDF analysis estimates the backgrounds from $W~+ \geq 2$~jets and $Z~+
\geq 2$~jets using the \vecbos~\cite{vecbos} event
generator followed by parton fragmentation using the 
\herwig~\cite{herwig} package and a Monte Carlo simulation of the CDF
detector.  The boson \pt\ requirement for \ww\ and \wz\ event
selection is chosen so that less than one background event is expected
in the final sample. Therefore, no background subtraction is necessary
and theoretical uncertainties in the background calculation are
avoided. Because no background subtraction is made, conservative
limits on anomalous couplings are obtained. In \d0\ the $W~+ \geq
2$~jets background is estimated with \vecbos, \herwig\, and a
\geant~\cite{geant} simulation of the \d0\ detector. The $W~+ \geq 2$~jets
background is normalized by comparing the number of events expected
from the \vecbos\ estimate to the number of candidate events observed
in the data outside the dijet mass window, after the multijet
background has been subtracted. Using this method the systematic
uncertainties in this background are due only to the normalization and
the jet energy scale uncertainty.

The data are in good agreement with the expected backgrounds plus SM
signal for both analyses. No excess of events at large $p_T^W$ is
observed and the overall shape of the $p_T^W$ distribution agrees well
with the predictions. Limits are derived using the leading-order
calculation by Hagiwara et al \cite{wwxs_lo} to
obtain the expected \ww\ and \wz\ signal as a function of the
anomalous couplings.

Varying only one coupling at a time and assuming equal \wwg\ and \wwz\
couplings, the 95\% CL limits obtained from
\d0\ and CDF are, for $\Lambda_{FF} = 2$~TeV, as follows:
\begin{tabbing}
{\d0}:~~~~~~~~~~~~\=$-0.43 < \Delta \kappa < 0.59$\\ 
		  \>$-0.33 < \lambda < 0.36$\\
CDF:              \>$-0.49 < \Delta \kappa  < 0.54$\\
                  \>$-0.35 < \lambda < 0.32$.
\end{tabbing}


Figure~\ref{fig:wwlvjj_limits} shows the limits obtained in the CDF
analysis.  Figure~\ref{fig:wwlvjj_limits}\textit{a} shows limits in the
$\lambda_\gamma - \lambda_z$ plane with all other couplings held at
their SM value.  The limits are stronger for $\lambda_z$, illustrating
the fact that this analysis is in general more sensitive to the \wwz\
coupling parameters (see {Section~\ref{sec:pheno}}).

The limits of Figure~\ref{fig:wwlvjj_limits}\textit{b} focus on the \wwz\
vertex, assuming that the \wwg\ couplings take the SM values. The
point $\kappa_Z = \lambda_Z = 0$ representing the minimal U(1)$_{\rm EM}$-only
coupling (corresponding to zero \wwz\ coupling) lies outside the
allowed region and is excluded at the 99\% CL by both experiments. This
is the first direct evidence for the existence of the \wwz\ coupling, and for
the destructive interference between the $s$-channel and $t$- or $u$-channel
diagrams which takes place in the SM.

In Figure~\ref{fig:wwlvjj_limits}\textit{c} the \wwz\ and \wwg\ coupling
parameters are assumed to be equal ($\kappa_Z = \kappa_\gamma$,
$\lambda_Z = \lambda_\gamma$), while in
Figure~\ref{fig:wwlvjj_limits}\textit{d} the HISZ relations are assumed.

\subsection{\it \d0\ Combined Analysis of \wwg\ and \wwz\ Couplings}

\d0\ has performed a simultaneous fit to the photon $p_T$
distribution in the $W\gamma$ data, the lepton $p_T$ distribution in
the $WW \to \ell \nu \ell^\prime \nu^\prime$ data, and the $p_T^{e
\nu}$ distribution in the $WW/WZ \to e \nu jj$ data~\cite{d0_1b_wwdilep}. 
Limits on the \wwg\ and \wwz\ coupling parameters are extracted from
the fit, taking care to account for correlations between the
uncertainties on the integrated luminosity, the selection efficiencies,
and the background estimates. The fit is performed using the
parameters $\Delta \kappa, \lambda$, and $g_1^Z$ and also using the
set $\alpha_{B\phi}, \alpha_{W\phi}$, and $\alpha_W$. The results are
given in Figures~\ref{fig:combined} and \ref{fig:combined-alpha} and 
Tables~\ref{table:combined} and \ref{table:combined-alpha}.
These are the most stringent limits to date on the \wwg\ and \wwz\ coupling
parameters $\Delta\kappa$, $\lambda$, and $\Delta g_1^Z$.

The \d0\ limits also provide the most stringent constraints on the 
parameters $\alpha_{B\phi}$ and $\alpha_W$.
The LEP measurements are more sensitive to $\alpha_{W\phi}$ than to 
$\alpha_{B\phi}$ and $\alpha_W$. 
The LEP limits are complimentary to the Tevatron limits because they are obtained
from a different process (i.e. $e^+ e^- \to W^+W^-$) using angular
distributions of the decay products.

\subsection{\it $Z \gamma$ Analysis Results}

\subsubsection{\footnotesize
$p \bar p \to \ell^+ \ell^- \gamma + X$}
This subsection describes the search for \zg\ events in which the
$Z$ decays to $e^+e^-$ or $\mu^+ \mu^-$. The event selection
requirements are similar to those for the \wg\ analysis except that
instead of the missing transverse energy requirement, a second charged
lepton is required with looser particle identification criteria.  The
photon selection requirements are almost identical to those used in
the \wg\ analyses as listed in Table~\ref{table:wg_evtsel}.  The
CDF analyses~\cite{cdf_wgzg1a_prd, cdf_wg1b,
cdf_zg1a_prl} and the \d0\ analyses~\cite{d0_diboson_1a_prd, d0_zg1a_prl, d0_zg1b_prl} are described elsewhere.

The main source of background is from $Z~+$~jet production where
the jet fakes a photon or an electron. The latter case corresponds
to the $e^+e^-$ signature if the track from one of the electrons from
the $Z \to e^+e^-$ decay is not reconstructed. A
smaller but nonnegligible background also resulting from particle
misidentification comes from QCD multijet and direct photon production,
where one or more jets are misidentified as electrons or photons.
These backgrounds are estimated from the number of $Z~+$~jet or
multijet/direct photon events observed in the data and the
misidentification probabilities $P(j \to ``\gamma$'') and $P(j \to
``e$''). The probabilities are obtained from multijet events as
described in {Section~\ref{sec:photon-detection}}.

The numbers of signal events after background subtraction are compared
with the SM predictions in Table~\ref{table:zg_nevts} for each experiment.
Figure~\ref{fig:zg_hists} shows kinematic distributions of the \d0\
candidate events together with the SM signal prediction plus the sum
of the estimated backgrounds. Two $ee\gamma$ events are observed with
photon $E_T \approx 75$~GeV and dielectron-photon invariant
mass $m_{ee\gamma} \approx 200$~GeV$/c^2$
(Figure~\ref{fig:zg_hists}\textit{a},\textit{c}). This is consistent with a fluctuation
of the SM signal. The probability of observing two or more events in
the combined electron and muon channels with $E_T^\gamma > 70$~GeV is
7.3\% for SM \zg\ production, and Monte Carlo studies show that the most
probable dielectron-photon invariant mass for events with $E_T^\gamma
= 70$--80~GeV is 200~GeV$/c^2$.  In both experiments, the numbers of
observed events and the shapes of the distributions show no deviations
from the expectations of the SM.

Limits on the anomalous coupling parameters are obtained using a
binned likelihood fit to the photon $E_T$ distribution as in the \wg\
analyses. The \zg\ signal prediction used is based on the leading-order calculation of Baur \& Berger~\cite{Baur&Berger-zgmc}. The
resulting 95\% CL limits on the CP-conserving \zzg\ and \zgg\ coupling
parameters are listed in Table~\ref{table:zg_limits}. Limits on the
CP-violating coupling parameters $h_{10}^V$ and $h_{20}^V$ are
numerically the same as the limits on $h_{30}^V$ and $h_{40}^V$.
Figure~\ref{fig:zg_limits_sum} shows the \d0\ limits in the $h_{30}^Z -
h_{40}^Z$ and $h_{30}^\gamma - h_{40}^\gamma$ planes.

\subsubsection{\footnotesize THE \d0\ ANALYSIS OF 
$p \bar p \to \etmis \gamma + X$}

\d0\ has carried out the first measurement of \zg\ 
production in the $Z \to \nu \bar \nu$ decay channel at a hadron
collider, and has demonstrated the higher sensitivity of
this channel to \zzg\ and \zgg\ anomalous couplings compared with the
channel $p \bar p \to \ell^+ \ell^- \gamma + X$~\cite{d0_diboson_1a_prd, d0_zvvg_prl}.
The neutrino decay channel has several 
experimental advantages over
the $\ell^+ \ell^- \gamma$ channel: the radiative decay background resulting
from the emission of a photon from the charged leptons in $Z \to
\ell^+ \ell^-$ decays is not present; the branching
fraction is higher,
$B (Z \to \nu \bar \nu) / B (Z \to \ell^+ \ell^-) \approx 3$, where
$\ell = e,\mu$;
and the efficiency is high since only one photon has to be detected as opposed
to a photon plus two charged leptons.

Although these factors result in a higher sensitivity to anomalous
couplings, there are also some disadvantages---there are additional
backgrounds, and the $Z$ boson cannot be identified since its
mass cannot be reconstructed.

There are two sources of instrumental backgrounds. One is due to cosmic
ray muons or beam halo muons that traverse the detector and emit a
photon by bremsstrahlung, which may deposit an energy cluster in the
EM calorimeter, as illustrated in Figure~\ref{fig:muon_brem}. 
If the muon is not reconstructed in the detector, the
resulting event signature is a single photon with balancing missing
\et.  The second source is due to $W \to e \nu$ events in which the electron
is misidentified as a photon, which occurs if the electron track is
not reconstructed in the central tracking chambers. There are also
physics backgrounds from QCD processes:
multijet production where a jet is misidentified as a photon and the
missing \et\ is due to mismeasured jets; 
direct photon production in which a jet contributes to \etmis; and 
$Z~+$~jets~$\to \nu \bar \nu~+$~jets events in which a jet is
misidentified as a photon.

To reduce the backgrounds from QCD processes and from $W \to e \nu$, tight
requirements are made on the photon \et\ and the missing transverse
energy: $E_T^\gamma > 40$~GeV and $\etmis > 40$~GeV.  These requirements
reduce the QCD backgrounds to negligible levels.  However, in
$W \to e \nu$ decays the electron \et\ distribution has a peak at $E_T
\approx 40$~GeV, and events in the tail of the Jacobian result in a
significant source of background. 

Two methods are used to further reduce this background.  The first
utilizes the fact that the Jacobian edge of the $E_T^e$ distribution
is smeared if the $W$s are produced with significant transverse
momentum due to initial-state radiation of gluons, illustrated
in Figure~\ref{fig:jacobian}. The number of electrons with $E_T^e >
40$~GeV is then higher relative to events with lower $p_T^W$. To
suppress the smearing of the Jacobian edge, thereby reducing the $W
\to e \nu$ background, a jet veto is applied, which rejects any event
containing a jet with $E_T^j > 15$~GeV. This method has the high
efficiency of 85\% for retaining $Z(\nu \bar \nu)\gamma$
events.

The second method, to reject electrons which do not have
reconstructed tracks, applies a cut on the number
of hits detected in each tracking chamber within a road defined between
the electromagnetic cluster's energy-weighted center and the event
vertex. The efficiency of this cut is approximately 75\%.
This technique gives a rejection factor of $r_h \approx 45$ 
and provides powerful background rejection
when combined with the rejection factor for the track match
requirement, which has $r_h \approx 5$.

The muon bremsstrahlung background is significantly suppressed by 
applying the following requirements:\\

\begin{enumerate}
\item The electromagnetic energy cluster must point back to
the interaction vertex. A
straight line $\chi^2$ fit is performed using
the energy-weighted centers of the EM cluster in all four layers of
the calorimeter plus the event vertex position. The resulting
probabilities $P_{xy}(\chi^2)$ and $P_{rz}(\chi^2)$ in the 
$xy$ and $rz$ planes are required to be
greater than 1\%. The vertex resolutions measured using $Z \to ee$
events are $\sigma_{xy}=11$~cm and $\sigma_{rz}=17$~cm and the
efficiency of the $P(\chi^2)$ requirement is 94\%.
\item No reconstructed muon is present in the CF muon chambers
($|\eta| < 1.0$). Typically, cosmic ray muons producing bremsstrahlung
photons consistent with the interaction vertex traverse the detector
in the central region. The efficiency of this requirement
is approximately 99\%;
\item No muon is identified by an energy
deposition in the finely segmented calorimeter, forming a track in a
road defined by the energy-weighted center of the EM cluster and the
interaction vertex. These events are predominantly from cosmic ray
and beam halo muon bremsstrahlung. This requirement has an efficiency
of 97\%.
\end{enumerate}

Applying all the requirements described above, the total estimated
background is $5.8 \pm 1.0$ events, with $4.0 \pm 0.8$ events from $W \to e
\nu$ and $1.8 \pm 0.6$ events from muon bremsstrahlung.  The expected
number of signal events for the SM and for anomalous couplings is
estimated using a leading-order \zg\ event
generator~\cite{Baur&Berger-zgmc} combined with the parametrized \d0
detector simulation described previously. For the SM the expected
number of signal events is $1.8 \pm 0.2$.  Four candidate events are
observed in the data, consistent with the SM expectations.

Limits on the anomalous couplings are derived from a maximum
likelihood fit to the photon \et\ spectrum, and are
listed in Table~\ref{table:zg_limits}. These limits are the most
stringent limits obtained from any one decay channel.

\section{\rm\large PROSPECTS FOR FUTURE STUDIES OF\\
ANOMALOUS COUPLINGS}

The analysis performed at the Tevatron will be repeated at
future machines, with increased energy and luminosity, where the
data will be much more sensitive to the virtual effects
that generate deviations from the SM expressions for the triple
gauge boson couplings. 

This section reviews the
expected sensitivity of experiments at LEPII, the Tevatron, the Large
Hadron Collider (LHC)~\cite{lhc.general}, and the Next Linear Collider
(NLC)~\cite{nlc.general}. Various options for
collision energies and integrated
luminosities have been considered for a linear collider.
We provide results for
representative cases at the 95\% CL (unless stated otherwise). 

The expected sensitivity of the NLC will be
sufficient to probe the SM radiative corrections 
(both electroweak and strong) to the processes
involving triple boson couplings; the
theoretical expectation for all contributions to the effective
parameters generated by non-SM physics will be subdominant and must be
extracted  as deviations from these radiative corrections.
We would also like to remark that, for any given process, there are
in principle a large number of terms in the effective Lagrangian
that generate deviations from the SM predictions. For example,
the process $ W^+ W^- \rightarrow W^+ W^- $ in the case where light
scalars are present is affected by the trilinear vertices involving 
gauge bosons, as well as by the scalar-$W$ couplings. Moreover, since
the initial-state $W$ bosons are radiated from a fermion, the process
is also affected by non-SM fermion-$W$ couplings. 

\subsection{\it LEP~II and the Tevatron}

The LEP~II experiments each collected $\sim 50$~pb$^{-1}$ in 1997 
at a CM energy of $183$~GeV. The
limits on anomalous couplings from these data are expected to be
a factor of about three better than the present LEP~II limits 
(see Table~\ref{table:combined-alpha}). If a total integrated
luminosity of 500~pb$^{-1}$ per experiment is achieved in the future,
the limits on the anomalous couplings will have a precision of 
$0.02-0.1$~\cite{LEPII-yellow}.

The expected integrated luminosity at the Tevatron in Run~II, 
which will start in the year 2000, is
$\approx 2-4$~fb$^{-1}$. Further upgrades in the accelerator complex
may result in data samples of up to 30~fb$^{-1}$. If 
10~fb$^{-1}$ is achieved, limits on 
anomalous couplings are expected to improve by a factor of 
about five~\cite{tev2000}.
With $\approx 2-4$~fb$^{-1}$ in Run~II, the Tevatron
also provides a unique opportunity to observe the SM radiation zero in 
$p \bar p \to W \gamma$.

\subsection{\it LHC}

Extracting deviations from the SM from
LHC data is complicated by the large contributions
generated by the QCD corrections~\cite{wwv.4,abraham}. The
expected limits from the reactions 
$ pp \rightarrow WZ,\,W\gamma $
for an integrated luminosity of 100 fb$^{-1}$
are~\cite{future.10,atlas}
\begin{eqnarray}
     | \Delta \kappa_Z | < 0.07 &&      | \lambda_Z | <0.005 \cr
| \Delta \kappa_\gamma | < 0.04 && | \lambda_\gamma | <0.0025
\label{lhc.limits}
\end{eqnarray}
[which differ from the results of Baur et al~\cite{future.11} due to the choice
of form factor scale $\Lambda_{FF}$. Fouchez~\cite{atlas} chose
$\Lambda_{FF}=10$~TeV which is much larger than the effective $\hat s$
of $\sim 1.4$~TeV; Baur et al took $\Lambda_{FF}=1$~or~3~TeV.]

These values will not be sufficient to
probe new physics at a scale above the effective CM
energy of the hard scattering process. Using the estimates
obtained in Section~{\ref{sec: coefficient.estimates}}, the above bounds
imply that the scale of new physics is larger than $ \sim 300$~GeV,
while the effective CM energy is $ \sim 1.2$~TeV~\cite{future.11}.
In other words, there are no models with a scale above $ 1.2 $~TeV
that produce deviations larger than those indicated in
Equation \ref{lhc.limits}.

Concerning the sensitivity to the neutral vector boson
vertices, the LHC is expected to achieve the limits~\cite{zgv.2}
\begin{equation}
\left| h_3^Z \right| < 2 \times 10^{-6} \qquad 
\left| h_4^Z \right| < 10^{-5}.
\end{equation}

\subsection{\it ep Collisions at the LHC}

This proposed collider, which  would collide protons in the LHC ring
with electrons in a reconstructed LEP ring
would be able to probe the trilinear
gauge boson vertices, but the bounds will not improve
on any obtained from LEP~II. The bound estimates for an integrated 
luminosity of 1 fb$^{-1}$ at 90\%~CL are given
in Table~{\ref{tab:lep.lhc}}~\cite{future.26}, where collisions of 
55~GeV electrons on 8~TeV protons were assumed.

This collider will not be able to probe
physics that cannot be directly produced for the 
integrated luminosity for which these studies were carried out.

\subsection{\it NLC}

The planned linear collider will be the first machine that can probe effective parameters at a level allowing
derivation of constraints on the scale of new physics superior to those
obtained from direct production.

Studies have been done for $e^+ e^- $, $ e\gamma $, and $ \gamma
\gamma $ initial states [the last two using back-scattered 
lasers~\cite{ginzburg}]. Although the CM energy of the machine has not
been definitely chosen, it is expected to operate
at $0.5$~TeV  for a first stage and then be upgraded to $1.5$~TeV.
There have been various studies of the sensitivity of these machines
to the effective couplings in 
Equation~\ref{wwv}~\cite{wwv.5prime,wwv.3.wwv.5.wwv.13.wwv.15.wwv.16,wwv.6,wwv.12}.
The most recent of these~\cite{wwv.6} makes a global 5 parameter 
fit to the coefficients $\Delta \kappa_{\gamma}$, $Delta \kappa_Z$,
$\lambda_\gamma$, $\lambda_Z$, and $ \Delta g_1^Z $ in
Equation~\ref{wwv}, which we reproduce in Table~{\ref{tab:nlc}}.
The table also includes limits for the couplings 
in Equation~\ref{zgv}~\cite{zgv.11.zgv.4} obtained using
various asymmetries.

The above sensitivity limits are strong enough to  insure that the NLC
will be able to probe new physics at scales beyond its CM energy. Although
this collider will also probe other reactions where new physics
effects can be significantly larger, the type of physics that modifies
the triple gauge boson vertices might not affect those other observables.

Alhough the above estimates give very tight limits, a complete 
multiparameter study including initial-state radiation effects and
detector efficiencies is still lacking.

\section{\rm \large SUMMARY}

We have reviewed studies of the trilinear gauge boson couplings
from the Tevatron Run~I data, with an integrated luminosity of
$\approx 100$~pb$^{-1}$.
Using gauge boson pair production processes, these measurements
provide the first direct tests of the trilinear gauge boson couplings. 

Limits on the \wwg\ effective couplings rule out the 
U(1)$_{\mathrm EM}$-only coupling of the $W$ boson to the photon 
($\kappa = \lambda = 0$) 
and also rule out a zero $W$ boson magnetic moment ($\mu_W =0$).
Studies of $WW$ and $WZ$ production are
also sensitive to the \wwz\ coupling and, for the first time, the
Tevatron measurements provide direct evidence for the existence 
of the \wwz\ coupling.

A simultaneous fit to the processes sensitive
to the \wwg\ and \wwz\ couplings provides the most stringent direct
limits to date on the effective couplings (for $\Lambda_{FF} = 2$~TeV
and assuming $\Delta\kappa_\gamma = \Delta\kappa_Z, 
\lambda_\gamma = \lambda_Z$):

\begin{center}
$-0.30 < \Delta \kappa < 0.43$\\
$-0.20 < \lambda < 0.20$\\
$-0.52 < \Delta g^Z_1 < 0.78$;\\
\end{center}

\noindent 
or, in the $\alpha_{B\phi}$, $\alpha_{W\phi}$, $\alpha_W$ parametrization:

\begin{center}
$-0.73 < \alpha_{B\phi} < 0.58$\\
$-0.22 < \alpha_{W\phi} < 0.44$\\
$-0.20 < \alpha_W < 0.20$.\\
\end{center}

Tests of the \zzg\ and \zgg\ effective couplings also provide the most
stringent limits to date. For a form factor scale $\Lambda_{FF} =
750$~GeV the limits are

\begin{center}
$\left| h^V_{10, 30} \right| < 0.36$\\
$\left| h^V_{20, 40} \right| < 0.05$.\\
\end{center}

While these measurements do not yet rule out any specific model beyond the
SM, the measurements are of crucial importance because they
test the trilinear gauge boson couplings, which are a fundamental
prediction of the SM, resulting from the non-Abelian
nature of the theory. 
It is worth pointing out that precision measurements
of this character have provided some of the most striking breakthroughs 
in particle physics---examples are
the anomalous magnetic moment of the electron and the Dirac theory, 
and $K^0$--$\overline{K}^0$ mixing and CP violation.

The typical values for effective couplings in
models beyond the SM are at the level $\leq 0.02$ 
(see Table~\ref{tab:table2}). Therefore, as the precision
improves in the future, experiments
will yield valuable information about new physics that
could give rise to anomalous couplings. The next measurements will
be made at 
LEP~II, the Tevatron, the LHC, and the planned NLC. Even if new physics is directly discovered, measurement of the loop
corrections to the triliner gauge boson couplings will still provide a
critical test of self-consistency of the theory.

\bigskip
\noindent
{\large {\bf ACKNOWLEDGMENTS}}\

\bigskip
We are indebted to the \d0\ and CDF Collaborations 
for making their results available to us. J Ellison
would like to thank his \d0\ colleagues who worked on the diboson
analyses, as well as Ulrich Baur for many useful, stimulating discussions.
We thank Greg Landsberg for providing the event displays used in
Figure~\ref{fig:muon_brem}, and Ann Heinson, who carefully read our
manuscript and made valuable suggestions.  This work was supported by
the US Department of Energy.

\begin{table}
\begin{center}
{\footnotesize
\begin{tabular}{cccccccc}
\hline \hline
& oblique & $ (g-2)_\mu $  & $d_n $  & $d_e $  & $ b \rightarrow s \gamma $
     & Atomic & $ K^0_L \rightarrow \mu \mu $ \\  

& params. & \cite{bound.28.bound.29.bound.48,zgv.29} &
\cite{wwv.5prime} & \cite{bound.54.bound.27.bound.1} &
\cite{bound8.bound10.bound49.bound59} & parity  &
\cite{bound.55}           \\ 

&\cite{bound.23} & & & & & viol.\cite{bound.55} & \\ \hline

\hline

$ | \Delta \kappa_\gamma | $  & $ 0.05 $                & $ 1 $                             & --                & --                               & $ 2 $                                 & $ 1 $                 & $ 1 $                     \\ 
$ | \Delta \kappa_Z | $       & $ 0.4 $                 & --                                & --                & --                               & --                                    & $ 0.12 $              & --                        \\ 
$ | \lambda_\gamma | $        & $ 0.2 $                 & $ 2 $                             & --                & --                               & $ 7 $                                 & $ 0.13 $              & --                        \\ 
$ | \lambda_Z | $             & $ 0.2 $                 & --                                & --                & --                               & --                                    & $ 0.13 $              & --                        \\ 
$ | \tilde \kappa_\gamma | $  & --                      & --                                & --                & $ 0.14 $                         & $ 0.4 $                               & --                    & --                        \\ 
$ | \tilde \kappa_Z | $       & --                      & --                                & --                & $ 0.04 $                         & --                                    & --                    & --                        \\ 
$ | \tilde \lambda_\gamma | $ & --                      & --                                & 0.00025           & --                               & $ 1.3 $                               & --                    & --                        \\ 
$ | g_4^Z | $                 & --                      & --                                & --                & $ 0.80 $                         & --                                    & --                    & --                        \\ 
$ | h_3^\gamma | $              & --                      & $ 4.5 $                           & --                & --                               & --                                    & --                    & --                        \\ \hline
\end{tabular}
}
\end{center}
\caption{
{\footnotesize
Indirect upper bounds on the effective 
parameters from precision measurements.}
}
\label{tab:table1}
\end{table}

\begin{table}
\begin{center}
{\footnotesize
\begin{tabular}{cccccc}
\hline \hline
Model          & $|\Delta\kappa_\gamma|$          & $|\lambda_\gamma|$              & $|\tilde\kappa_\gamma|$             \\ \hline
\sm            & $0.008$~\cite{sm19,sm6}          & $0.002$~\cite{sm6}              & $10^{-22}$~\cite{sm13,sm20}         \\
2HDM           & $0.016$~\cite{2hdm5}             & $0.0014$~\cite{2hdm5}           & --                                  \\
Multi-doublet  & --                               & --                              & $4 \times 10^{-6}$~\cite{sm12,sm13} \\         
E6             & $2.5 \times 10^{-5}$~\cite{sm11} & $0.003$~\cite{sm11}             & --                                  \\
SUSY           & $0.005$~\cite{susy2}             & $5 \times 10^{-5}$~\cite{susy2} & $3 \times 10^{-4}$~\cite{susy7}     \\
TC             & $0.002$~\cite{tc2}               & --                              & $7 \times 10^{-6}$~\cite{tc2}       \\
4th generation & --                               & --
& $5 \times 10^{-3}$~\cite{various10} \\ \hline
\end{tabular}
}
\end{center}
\caption{
{\footnotesize
Calculated values of the effective parameters in several theoretical
models. The abbreviations are 2HDM (two Higgs-doublet model), 
SUSY (Suppersymmetry), and TC (Technicolor).
}
}
\label{tab:table2}
\end{table}

\begin{table*}[ht]
\begin{center}
{\footnotesize 
\begin{tabular}{lcccc} \hline\hline
    & \multicolumn{2}{c} { \D0 }
    & \multicolumn{2}{c}{ CDF }  \\
    & $W\gamma\rightarrow e\nu\gamma $  
    & $W\gamma\rightarrow \mu\nu\gamma $
    & $W\gamma\rightarrow e\nu\gamma $  
    & $W\gamma\rightarrow \mu\nu\gamma $
\\ \hline
Lepton $\eta$      & $|\eta_e| < 1.1$~or     & $|\eta_\mu| < 1.0$ 
                   & $|\eta_e| < 1.1$        & $|\eta_\mu| < 0.6$    \\
	           & $1.5 < |\eta_e| < 2.5$  & 
		   &			     & \\

Lepton $p_T$ (GeV$/c$)   & $E_T^e > 25$            & $p_T^\mu > 15$ 
                         & $E_T^e > 20$            & $p_T^\mu > 20$      \\

Missing $E_T$     (GeV)  & $\etmis > 25$          & $\etmis > 15$
                         & $\etmis > 20$          & $\etmis > 20$      \\

Photon $\eta$          & \multicolumn{2}{c} {$|\eta_\gamma| < 1.1$~or~$1.5 < |\eta_\gamma| < 2.5$ }
		       & \multicolumn{2}{c} {$|\eta_\gamma| < 1.1$} \\

Photon $E_T$     (GeV)      & \multicolumn{2}{c} {$E_T^\gamma > 10$}
			    & \multicolumn{2}{c} {$E_T^\gamma > 7$} \\

$\ell-\gamma$ separation    & \multicolumn{2}{c} {$\Delta
			    R_{\ell \gamma} > 0.7$}
			    & \multicolumn{2}{c} {$\Delta
			    R_{\ell \gamma} > 0.7$} \\
\hline
\end{tabular}
}
\end{center}
\caption{
{\footnotesize
Summary of \wg\ event selection requirements}
}
\label{table:wg_evtsel}
\end{table*}
%

%
\begin{table*}[ht]
\begin{center}
{\footnotesize 
\begin{tabular}{lcccc} \hline\hline
    & \multicolumn{2}{c} { \D0 }
    & \multicolumn{2}{c}{ CDF }  \\ 
    & \multicolumn{2}{c} { 92.8 pb$^{-1}$ }
    & \multicolumn{2}{c}{ 67.0 pb$^{-1}$ }  \\
    & $W\gamma\rightarrow e\nu\gamma $  
    & $W\gamma\rightarrow \mu\nu\gamma $
    & $W\gamma\rightarrow e\nu\gamma $  
    & $W\gamma\rightarrow \mu\nu\gamma $
\\ \hline
$N_{\rm data}$  & 57                      & 70 
                & 75                      & 34                  \\
$N_{\rm bkg}$   & 15.2 $\pm$ 2.5          & 27.7 $\pm$ 4.7 
                & 16.1 $\pm$ 2.4          & 10.3 $\pm$ 1.2      \\
$N_{\rm sig}$   & $41.8^{+8.8}_{-7.5}$    & $42.3^{+9.7}_{-8.3}$  
                & $58.9 \pm 9.0 \pm 2.6 $ & $23.7 \pm 5.9 \pm 1.1 $     \\
$N_{\rm SM}$    & 43.6 $\pm$ 3.1          & 38.2 $\pm$ 2.8 
                & $53.5 \pm 6.8$          & $21.8 \pm 4.3  $     \\
\hline
\end{tabular}
}
\end{center}
\caption{
{\footnotesize
Number of candidate \wg\ events observed
$N_{\rm data}$. $N_{\rm bkg}$, estimated
background; $N_{\rm sig}$, number of signal events after
background subtraction; $N_{\rm SM}$, SM prediction}
}
\label{table:wg_nevts}
\end{table*}
\begin{table*}[ht]
\begin{center}
{\footnotesize 
\begin{tabular}{lcccccc} \hline\hline
    & \multicolumn{3}{c} { \D0 }
    & \multicolumn{3}{c}{ CDF }  \\
    & $e \nu e \nu$  
    & $e \nu \mu \nu$
    & $\mu \nu \mu \nu$   
    & \multicolumn{3}{c} {(all modes)}
\\ \hline
Electron $\eta$
& \multicolumn{3}{c} {$|\eta_e| < 1.1$~or}
& \multicolumn{3}{c} {$|\eta_e| < 1.0$~or} \\
& \multicolumn{3}{c} {$1.5 < |\eta_e| < 2.5$}
& \multicolumn{3}{c} {$1.20 < |\eta_e| < 1.35$} \\

Muon $\eta$
& \multicolumn{3}{c} {$|\eta_\mu| < 1.0$} 
& \multicolumn{3}{c} {$|\eta_\mu| < 1.2$} \\

Lepton $p_T$ (GeV$/c$)	& $E_T^{e1} > 25$
			& $E_T^e > 25$
			& $p_T^{\mu 1} > 25$
                     	& \multicolumn{3}{c} {$p_T > 20$} \\ 

		   	& $E_T^{e2} > 20$
			& $p_T^\mu > 15$
			& $p_T^{\mu 2} > 20$
            	    	& \multicolumn{3}{c} {(for all $e$'s, $\mu$'s)} \\ 

Missing $E_t$ (GeV)  & $\etmis > 25$
			 & $\etmis > 20$
                         & $-$
			 & \multicolumn{3}{c} {$\etmis > 20$} \\

$t \bar t$~rejection & \multicolumn{3}{c} {$E_T^{\rm had} < 40$~GeV}
		     & \multicolumn{3}{c} {no jets with} \\
 		     & \multicolumn{3}{c} {}
		     & \multicolumn{3}{c} {$E_T > 10$~GeV} \\

$Z$~rejection 	     & \multicolumn{3}{c} {see text}
		     & \multicolumn{3}{c} {see text} \\

\hline
\end{tabular}
}
\end{center}
\caption{
{\footnotesize
Summary of $WW \to \ell \nu \ell^\prime \nu^\prime$ event
selection requirements.}
}
\label{table:wwdilep_evtsel}
\end{table*}

\begin{table*}[ht]
\begin{center}
{\footnotesize 
\begin{tabular}{lcc} \hline\hline
    & { \d0\ }
    & { CDF }  \\
\hline
Electron $\eta$    & $|\eta_e| < 1.1$~or     & $|\eta_e| < 1.1$~or \\
	           & $1.5 < |\eta_e| < 2.5$  & $1.1 < |\eta_e| < 2.4$ \\

Muon $\eta$        &  --     & $|\eta_\mu| < 1.0$ \\

Lepton $E_T$ or $p_T$ (GeV)    & $E_T^e > 25$      & $E_T^e$, $p_T^\mu > 20$ \\

Missing $E_t$ (GeV)            & $\etmis > 25$     & $\etmis > 20$  \\

Transverse mass (GeV$/c^2$)    & $m_T(\ell;\etmis) > 40$   
                               & $m_T(\ell;\etmis) > 40$ \\

Jet cone radius   & $R = 0.3$~(Ia), 0.5~(Ib)   & $R = 0.4$ \\

Jet $\eta$	             & $|\eta_j| < 2.5$  & $|\eta_j| < 2.5$ \\

Jet $E_T$		     & $E_T^j > 20$      & $E_T^j > 30$ \\

Dijet invariant mass (GeV$/c^2$)  & $50 < m_{jj} < 110$ 
				  & $60 < m_{jj} < 110$  \\

Dijet \pt\ (GeV$/c$) 		& -- & $p_T^{jj} > 200$  \\	  

\hline
\end{tabular}
}
\end{center}
\caption{
{\footnotesize
Summary of $WW/WZ \to \ell \nu jj$  event selection requirements}
}
\label{table:wwlvjj_evtsel}
\end{table*}

\begin{table*}[ht]
\begin{center}
{\footnotesize
\begin{tabular}{ccc}  \hline \hline
Coupling	&$\Lambda_{FF}=1.5$~TeV	&$\Lambda_{FF}=2.0$~TeV  \\ \hline

$\Delta\kappa_{\gamma} = \Delta\kappa_Z$   &$-$0.33, 0.46   &$-$0.30, 0.43\\ 

$\lambda_{\gamma} = \lambda_Z$		   &$-$0.21, 0.21   &$-$0.20, 0.20\\

$\Delta\kappa_{\gamma}$ (HISZ)		   &$-$0.39, 0.61   &$-$0.37, 0.56\\

$\lambda_{\gamma}$ (HISZ)		   &$-$0.21, 0.21   &$-$0.20, 0.20\\

$\Delta g^Z_1$ (SM \wwg)		   &$-$0.56, 0.86   &$-$0.52, 0.78\\

$\Delta\kappa_Z$ (SM \wwg)		   &$-$0.46, 0.64   &$-$0.42, 0.59\\

$\lambda_Z$ (SM \wwg)			   &$-$0.33, 0.37   &$-$0.31, 0.34\\

$\Delta\kappa_{\gamma}$ (SM \wwz)	   &$-$0.63, 0.75   &$-$0.59, 0.72\\

$\lambda_{\gamma}$ (SM \wwz)	  	   &$-$0.27, 0.25   &$-$0.26, 0.24\\
\hline 
\end{tabular}
}
\end{center}
\caption{
{\footnotesize
\d0\ limits on anomalous couplings at the 95\% CL from a 
   simultaneous fit to the $W\gamma$, $WW \to \ell \nu \ell^\prime
   \nu^\prime$, and $WW/WZ \to e \nu jj$ data.}
}\label{table:combined}
\end{table*}
\begin{table*}[ht]
\begin{center}
{\footnotesize
\begin{tabular}{cccc}  \hline \hline
Coupling   &$\Lambda_{FF}=1.5$~TeV   &$\Lambda_{FF}=2.0$~TeV  & LEP combined \\ \hline

$\alpha_{B\phi}$  & $-$0.76, 0.61  & $-$0.73, 0.58  & $-$0.81, 1.50\\

$\alpha_{W\phi}$  & $-$0.24, 0.46  & $-$0.22, 0.44  & $-$0.28, 0.33\\

$\alpha_W$	  & $-$0.21, 0.21  & $-$0.20, 0.20  & $-$0.37, 0.68\\

$\Delta g^Z_1$	  & $-$0.31, 0.60  & $-$0.29, 0.57  & $-$\\
\hline
\end{tabular}
}
\end{center}
\caption{
{\footnotesize
\d0\ limits on anomalous couplings $\alpha_{B\phi}$, 
$\alpha_{W\phi}$, $\alpha_W$, and $\Delta g^Z_1$
at the 95\% CL from a simultaneous fit to 
the $W\gamma$, $WW \to \ell \nu \ell^\prime \nu^\prime$, 
and $WW/WZ \to e \nu jj$ data. The last column gives the combined
limits from the LEP experiments \cite{lep-tgc-results}.
The LEP limits should be multiplied by a factor 
$(1 + s / \Lambda^2_{FF})^2$ to compare directly with the
\d0\ results. At the LEP energy $\sqrt{s} = 172$~GeV, 
this factor is only 1.026 for $\Lambda_{FF} =  1.5$~TeV so the numbers
in the table have not been corrected.}
}
\label{table:combined-alpha}
\end{table*}

\begin{table*}[ht]
\begin{center}
{\footnotesize 
\begin{tabular}{lcccc} \hline\hline
    & \multicolumn{2}{c} { \D0 }
    & \multicolumn{2}{c}{ CDF }  \\ 
    & $Z\gamma\rightarrow ee\gamma $  
    & $Z\gamma\rightarrow \mu\mu\gamma $
    & $Z\gamma\rightarrow ee\gamma $  
    & $Z\gamma\rightarrow \mu\mu\gamma $ 

\\ \hline

$\int {\cal L} dt$ (pb$^{-1})$  
		& 111.3	    & 100.7	  & 67.0    & 67.0	\\

$N_{\rm data}$  & 18                      & 17 
                & 18                      & 13                   \\

$N_{\rm bkg}$   & $2.24 \pm 0.54$         & $3.62 \pm 0.68$   
                & $0.9 \pm 0.3$           & $0.5 \pm 0.1$        \\

$N_{\rm sig}$   & $15.8^{+4.9}_{-4.3}$    & $13.4^{+4.7}_{-4.1}$ 
                & $17.1 \pm 5.7$    	  & $12.5 \pm 3.6$       \\

$N_{\rm SM}$    & $16.0 \pm 1.3$          & $18.6 \pm 2.0$ 
                & $16.2 \pm 1.8$          & $8.7 \pm 0.7$        \\
\hline
\end{tabular}
}
\end{center}
\caption{
{\footnotesize
$N_{\rm data}$, the number of candidate \zg\ events observed; $N_{\rm bkg}$, the estimated
background; $N_{\rm sig}$, the number of signal events after
background subtraction; $N_{\rm SM}$, the SM prediction.}
}
\label{table:zg_nevts}
\end{table*}

\begin{table*}[ht]
\begin{center}
{\footnotesize
\begin{tabular}{lcccc}   \hline\hline

Analysis 	& $\int {\cal L} dt$ (pb$^{-1}$) & $\Lambda_{FF}$ (GeV)
		& $h^Z_{30}$			 & $h^Z_{40}$ \\ \hline

CDF Ia $\ell^+ \ell^- \gamma$
	& 20	& 500	& $-$3.0, 2.9 	& $-$0.7, 0.7 \\

CDF Ia + Ib prelim.
	& 67 	& 500 	& $-$1.6, 1.6	& $-$0.4, 0.4 \\

\d0\ Ia $\ell^+ \ell^- \gamma$
	& 14 	& 500	& $-$1.8, 1.8	& $-$0.5, 0.5\\

\d0\ Ia $\nu \bar \nu \gamma$
	& 13	& 500	& $-$0.87, 0.87	& $-$0.19, 0.19\\
\hspace{0.3in} ''	& ''	& 750	& $-$0.49, 0.49	& $-$0.07, 0.07\\

\d0\ Ib $\ell^+ \ell^- \gamma$
	& 97 ($e$), 87($\mu$) 	& 500	& $-$1.31, 1.31	& $-$0.26, 0.26\\
\hspace{0.3in} ''	& ''	& 750	& $-$0.67, 0.67	& $-$0.08, 0.08\\

\d0\ combined
	& --	& 750	& $-$0.36, 0.36	& $-$0.05, 0.05\\

\hline 
\end{tabular}
}
\end{center}
\caption{
{\footnotesize
Summary of the 95\% CL limits on anomalous couplings from the
\zg\ analyses. All couplings other than those
indicated are held to their SM values. Limits on the
CP-violating coupling parameters $h_{10}^V$ and $h_{20}^V$ are
numerically the same as the limits on $h_{30}^V$ and $h_{40}^V$.}
Limits on the couplings $h_{10}^\gamma - h_{40}^\gamma$ are almost the
same as those on $h_{10}^Z - h_{40}^Z$. 
}
\label{table:zg_limits}
\end{table*}
\begin{table}
\begin{center}
{\footnotesize
\begin{tabular}{ c c c c }
\hline \hline
\multicolumn{4}{c}{LEP$\times$LHC expected limits}\\
\hline
$ | \lambda_\gamma           | < 0.12  $ &
$ | \lambda_Z                | < 0.3   $ &
$ | \tilde \lambda_\gamma    | < 0.12  $ & \\
$ | \tilde \lambda_Z         | < 0.24  $ & 
$ | \Delta \kappa_\gamma     | < 0.4   $ & 
$ | \Delta \kappa_Z          | < 0.8   $ & 
$ | \tilde \kappa_Z          | < 1     $   \\
$ | g_4^Z                    | < 0.35  $ &
$ | g_5^Z                    | < 0.5   $ &
\multicolumn{2}{l}{$-0.6 < g_1^Z < 0.3$}   \\
\hline
\end{tabular}
}
\end{center}
\caption{
{\footnotesize
Expected limits on effective couplings at the 90\% CL for
the LHC$\times$LEP collider with an integrated luminosity of
1 fb$^{-1}$. The reactions studied are $ e^- p \rightarrow e^\pm
W^\mp X$, and $\nu Z X$}
}
\label{tab:lep.lhc}
\end{table}
\begin{table}
\begin{center}
{\footnotesize
\begin{tabular}{r l l c r l l  }
\hline \hline
$\sqrt{s}$:     &0.5 TeV&1 TeV&\multicolumn{1}{c}{}&&0.5 TeV& 1 TeV  \\
\hline
$|\Delta\kappa_\gamma|$ & 0.002 & 0.0006 &&$|g_4^Z     |$ & 0.075 & 0.0024 \\
$|\Delta\kappa_Z     |$ & 0.05  & 0.02   &&$|h_1^\gamma|$ & 0.003 & --     \\
$|\lambda_\gamma     |$ & 0.011 & 0.003  &&$|h_1^Z     |$ & 0.02  & --     \\
$|\lambda_Z          |$ & 0.025 & 0.006  &&$|h_3^V     |$ & 0.03  & 0.005  \\
$|\Delta g_1^Z       |$ & 0.05  & 0.015  &&$|h_4^V     |$ & 0.003 & 0.0002 \\
\hline
\end{tabular}
}
\end{center}
\caption{
{\footnotesize
Expected limits on the effective couplings from the
initial and intermediate stages of the NLC.
Luminosities for the \wwg\ and \wwz\ coupling parameters
are 20 and 50 fb$^{-1}$ for 0.5 and 1~TeV CM energies
respectively. Limits
on the C odd, P even parameter
$g_4^Z$~\cite{wwv.4} are derived from
the asymmetries in 
$ e^+ e^- \rightarrow \nu \bar\nu Z $~\cite{wwv.12}, 
for luminosities of 
10 fb$^{-1}$ and 100 fb$^{-1}$. They hold for energies in the range 
0.3--2~TeV.}
}
\label{tab:nlc}
\end{table}

\clearpage
\begin{figure}[p]
    \epsfysize = 3cm
    \centerline{\epsffile{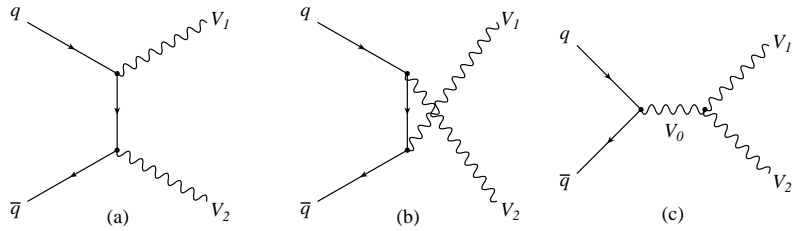}}
\caption{
{\footnotesize
Leading-order Feynman diagrams for vector boson pair
production. The assignments of $V_0$, $V_1$, and $V_2$ are: 
$V_0=V_1=W$ and $V_2=\gamma$ for \wg\ production; 
$V_0=\gamma$~or~$Z$, $V_1=W^+$ and $V_2=W^-$ for \ww\ production;
$V_0=V_1=W$ and $V_2=Z$ for \wz\ production; and 
$V_0=\gamma$~or~$Z$, $V_1=Z$ and $V_2=\gamma$ for \zg\ production.
}}
\label{fig:feynman1}
\end{figure}
\begin{figure}[p]
   \epsfysize = 2cm
    \centerline{\epsffile{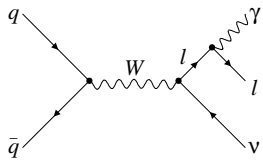}}
\caption{
{\footnotesize
Leading-order Feynman diagram for $W$ production with
radiative $W$ decay: The
charged lepton radiates a photon by bremsstrahlung.}
}
\label{fig:feynman2}
\end{figure}
\begin{figure}[p]
   \epsfysize = 6cm
   \centerline{\epsffile{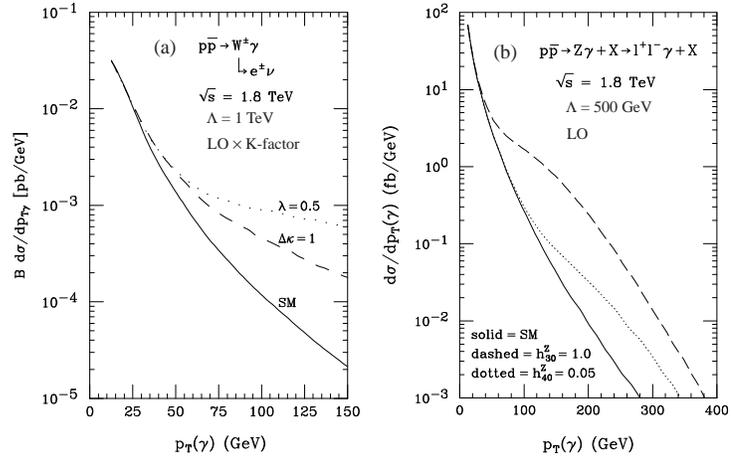}}
\caption{
{\footnotesize
Photon transverse momentum spectra for (\textit{a}) \wg\ production and (\textit{b})
\zg\ production at the Tevatron for SM and
anomalous couplings. From Refs.~\cite{Baur&Berger-wgmc} and
\cite{zg-nlo}.}
}
\label{fig:ptgamma}
\end{figure}

\clearpage

%
\begin{figure}[p]
   \epsfysize = 3.7cm
   \centerline{\epsffile{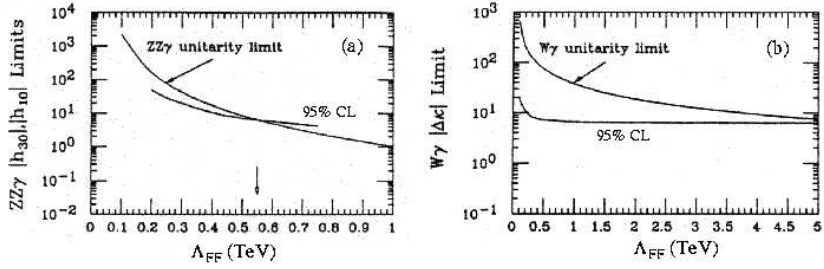}}
\caption{
{\footnotesize
Experimental 95\% confidence level limits and
unitarity limits as a function of form factor scale $\Lambda_{FF}$ for
(\textit{a}) the \zzg\ coupling $h^Z_{30}$ and (\textit{b}) the \wwg\ coupling
$\Delta\kappa$ [from ~\cite{cdf_wgzg1a_prd}].
Note that the experimental limits have been superseded by the much
tighter limits described in {Section~\ref{sec:analysis}}.}
}
\label{fig:ff-scale}
\end{figure}
\clearpage

\begin{figure}[p]
    \epsfysize = 7cm
    \centerline{\epsffile{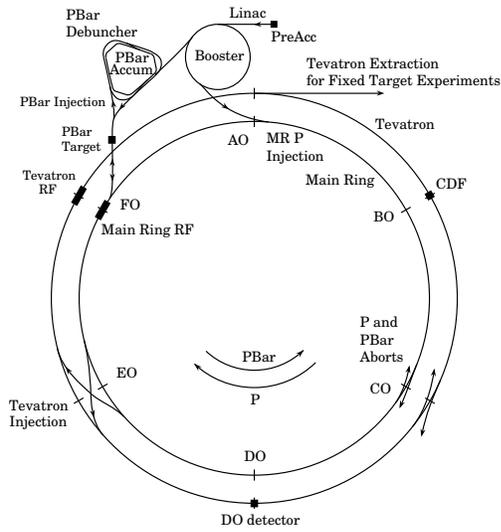}}
\caption{
{\footnotesize
The Fermilab accelerator complex, from Ref.~\cite{accelerators}.}
}
\label{fig:Tevatron}
\end{figure}

\begin{figure}[p]
    \epsfysize = 5cm
    \centerline{\epsffile{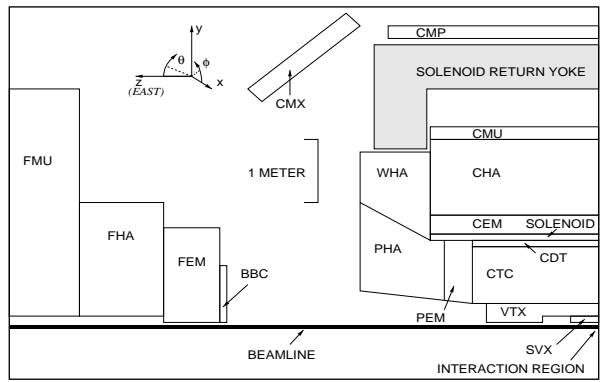}}
\caption{
{\footnotesize
A cross-sectional side view of the CDF detector 
[from~\cite{CDF-topPRD}]. SVX, VTX, CTC, and CDT are tracking
detectors. The calorimeters are CEM, CHA, WHA, PEM, PHA, FEM, and
FHA. The muon detectors are CMU, CMP, CMX, and FMU.}
}
\label{fig:CDF-detector}
\end{figure}
\clearpage

\begin{figure}[p]
    \epsfysize = 6cm
    \centerline{\epsffile{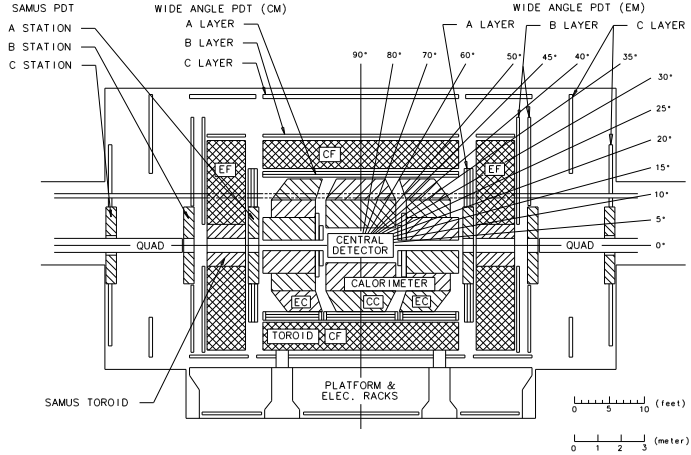}}
\caption
{\footnotesize
{Cross-sectional side view of the \d0\ detector [from~\cite{D0-detector}], showing the central detector, the
calorimeters (CC, EC), and the muon system (CF, EF, SAMUS, and
proportional drift chambers (PDTs).}
}
\label{fig:D0-detector}
\end{figure}

\begin{figure}[p]
    \epsfysize = 7cm
    \centerline{\epsffile{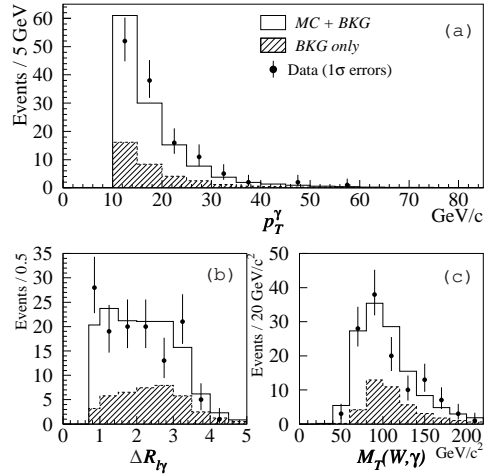}}
\vskip -5mm
\caption{
{\footnotesize
Distributions of (\textit{a}) photon transverse energy $p_T^\gamma$,
(\textit{b}) lepton-photon separation $\Delta R_{\ell \gamma}$, and (\textit{c}) transverse
cluster mass $M_T(W,\gamma)$ for the \d0\ \wg\ analysis.}
}
\label{fig:wghists_d0}
\end{figure}

\begin{figure}[p]
    \epsfysize = 3.5cm
    \centerline{\epsffile{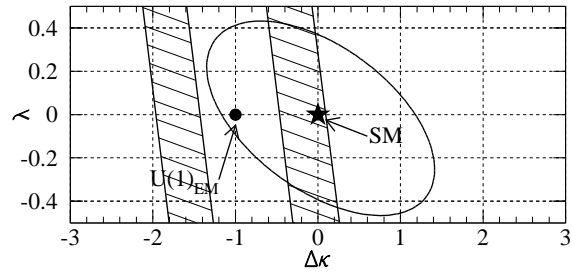}}
\caption[]{
{\footnotesize
Limits on the \wwg\ couplings at the 95\% confidence level (CL) for $\Lambda_{FF} = 1.5$~TeV
from \d0\ (\textit{ellipse}). The \textit{shaded bands} are the regions allowed by the
95\% CL limits from the CLEO and ALEPH observations of $b \to s
\gamma$ decays~\cite{bsgamma}.}
}
\label{fig:wglimits}
\end{figure}
\clearpage

\begin{figure}[p]
    \epsfysize = 5cm
    \centerline{\epsffile{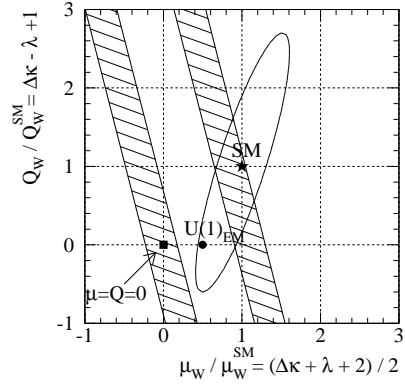}}
\caption[]{
{\footnotesize
Limits on the $W$ boson magnetic dipole moment $\mu_W$ and 
        electric quadrupole moment $Q_W$ at the 95\% confidence level from \d0\ 
        (\textit{ellipse}). The \textit{shaded bands} are the limits from
        $b \to s \gamma$~\cite{bsgamma}.}
}
\label{fig:qmulimits}
\end{figure}

\begin{figure}[p]
    \epsfxsize = 10.75cm
    \centerline{\epsffile{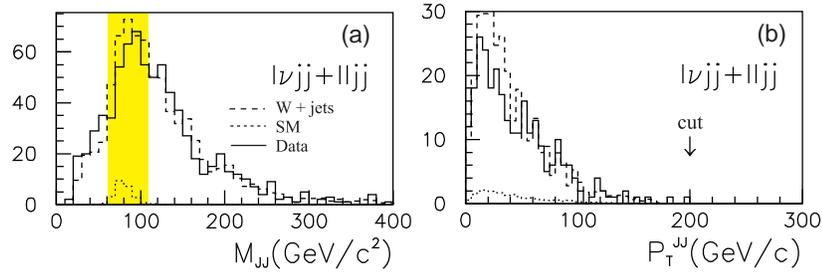}}
\caption{
{\footnotesize
CDF selection of $W/WZ \to \ell \nu jj, \ell \ell jj$ events.
(\textit{a}) Dijet mass distribution for events passing all selection requirements
except the dijet mass cut. (\textit{b}) The \pt\ of the two-jet system for the 
subset of events from (\textit{a}) passing the dijet mass cut [\textit{shaded
region} in (\textit{a})]. The
distributions are shown for the data (\textit{solid line}), the $W +$~jets
background (\textit{dashed line}), and the SM prediction (\textit{dotted line}).}  
}
\label{fig:wwlvjj_cdfhists}
\end{figure}
\begin{figure}[p]
    \epsfysize = 8.5cm
    \centerline{\epsffile{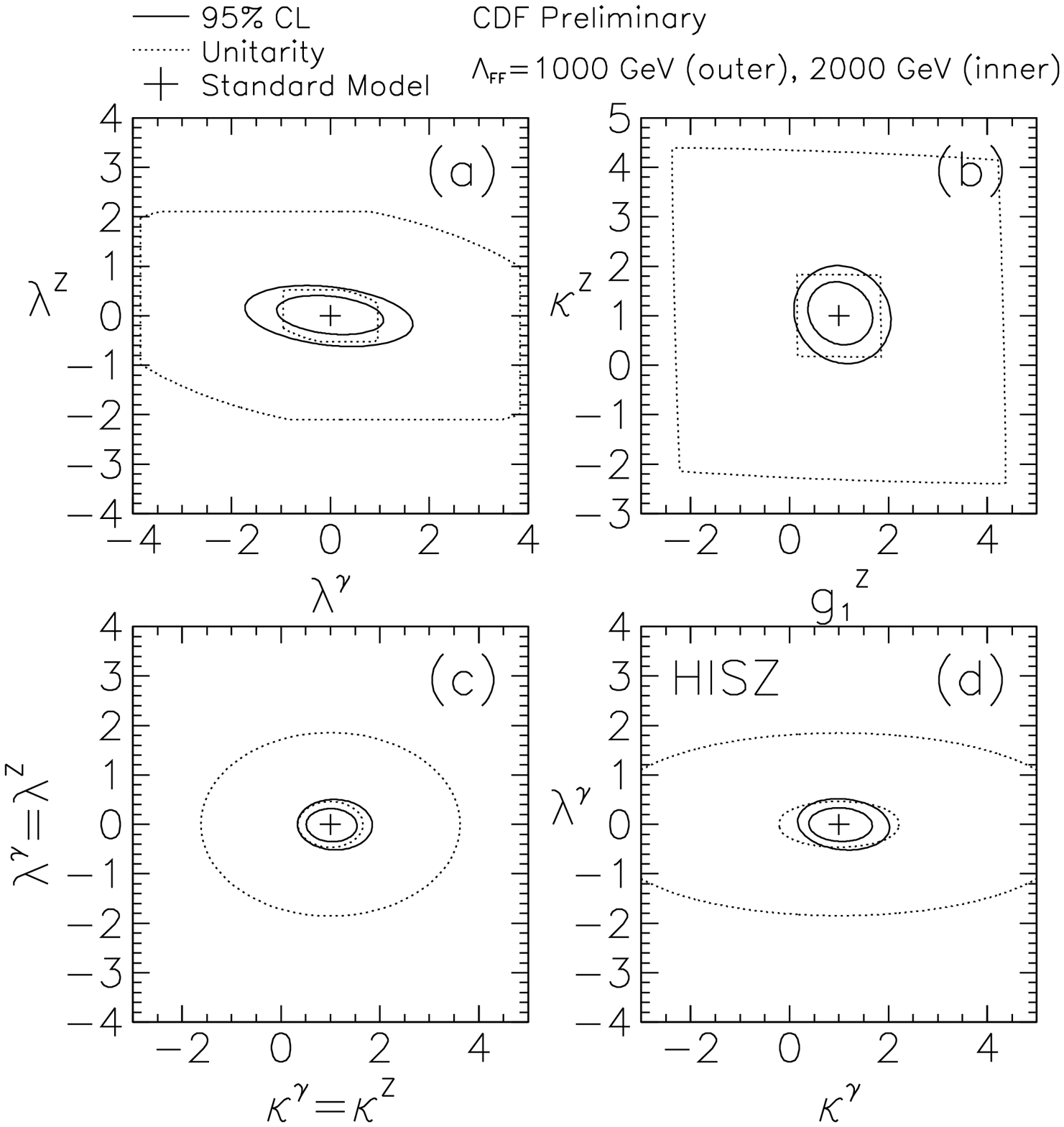}}
\caption{
{\footnotesize
Allowed regions for anomalous couplings from the preliminary
CDF analysis. All couplings, other than those plotted for each
contour, are held at their SM values: (\textit{a}) $\lambda_Z$
vs $\lambda_\gamma$; (\textit{b}) $\kappa_Z$ vs. $g_1^Z$; (\textit{c}) $\lambda$
vs $\kappa$ assuming the \wwz\ and \wwg\ coupling parameters are
equal; and (\textit{d}) limits on the couplings $\kappa_\gamma$, $\kappa_Z$,
$\lambda_\gamma$, $\lambda_Z$, and $g_1^Z$ in the HISZ scenario
with independent variables $\kappa_\gamma$ and $\lambda_\gamma$.}  
}
\label{fig:wwlvjj_limits}
\end{figure}

\begin{figure}[p]
    \epsfysize = 7cm
    \centerline{\epsffile{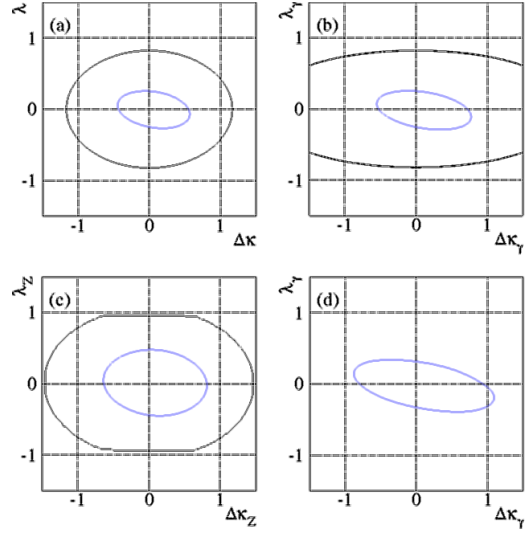}}
\caption{
{\footnotesize
\d0\ limits on anomalous couplings for
        $\Lambda_{FF} = 1.5$~TeV from a simultaneous fit to the $W\gamma$, 
	$WW \to \ell \nu \ell^\prime \nu^\prime$, and 
	$WW/WZ \to e \nu jj$ data: (\textit{a}) assuming $\Delta \kappa_Z = 
	\Delta \kappa_\gamma$, $\lambda_Z = \lambda_\gamma$; 
	(\textit{b}) assuming the HISZ scenario; (\textit{c}) assuming SM \wwg\
        couplings; and (\textit{d}) assuming SM \wwz\
        couplings. The inner curves are the 95\% CL limits and the outer curves
        are the unitarity limits. In (\textit{d}) the unitarity limits lie
        outside the boundary of the plot.}  
}
\label{fig:combined}
\end{figure}
\begin{figure}[p]
    \epsfysize = 3.25cm
    \centerline{\epsffile{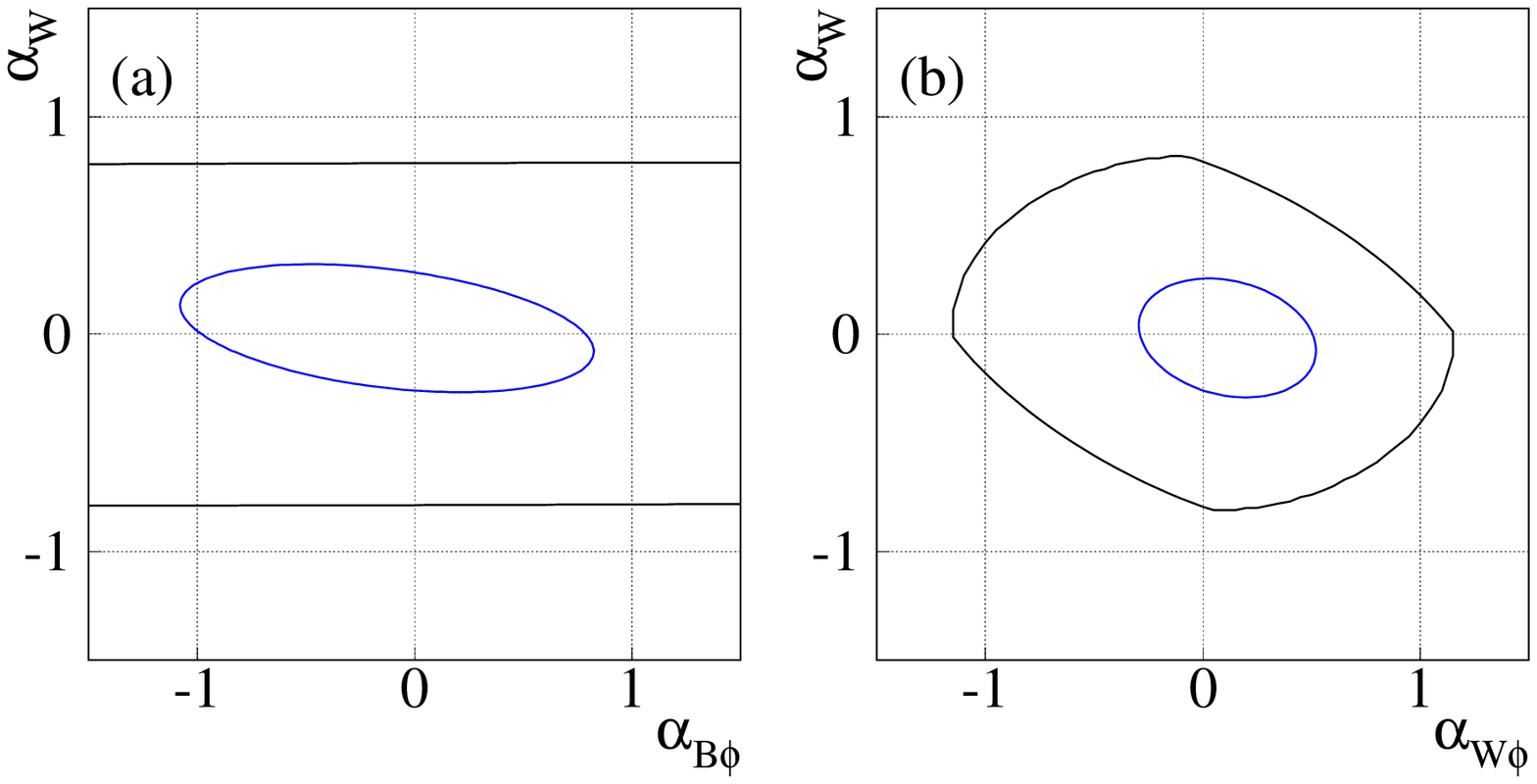}}
\caption{
{\footnotesize
\d0\ limits on anomalous couplings for
        $\Lambda_{FF} = 1.5$~TeV from a simultaneous fit to the $W\gamma$, 
	$WW \to \ell \nu \ell^\prime \nu^\prime$, and 
	$WW/WZ \to e \nu jj$ data for: (\textit{a}) $\alpha_W$ vs $\alpha_{B\phi}$
        when $\alpha_{W\phi}=0$; and (\textit{b})  $\alpha_W$ vs. $\alpha_{W\phi}$
        when $\alpha_{B\phi}=0$. The inner curves are the 95\% CL
        limits and the outer curves are the unitarity limits.}  
}
\label{fig:combined-alpha}
\end{figure}

\begin{figure}[p]
    \epsfysize = 7cm
    \centerline{\epsffile{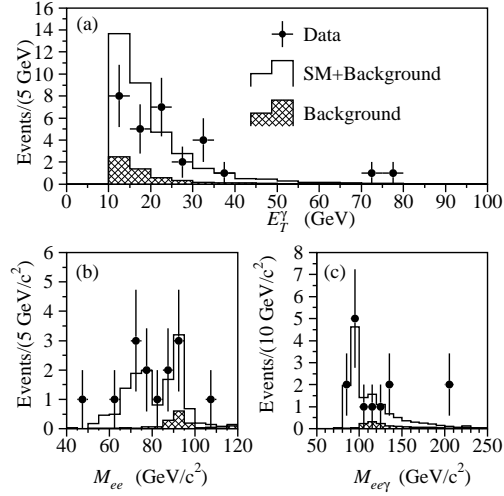}}
\caption{
{\footnotesize
Kinematic properties of the candidate events and estimated
backgrounds in the \d0\ $Z(\ell^+ \ell^-)\gamma$ analysis: (\textit{a}) photon
tranverse energy for the combined $ee\gamma$ and $\mu \mu \gamma$
samples; (\textit{b}) dielectron invariant mass; (\textit{c}) dielectron-photon
invariant mass.}
}  
\label{fig:zg_hists}
\end{figure}

\begin{figure}[p]
    \epsfysize = 7cm
    \centerline{\epsffile{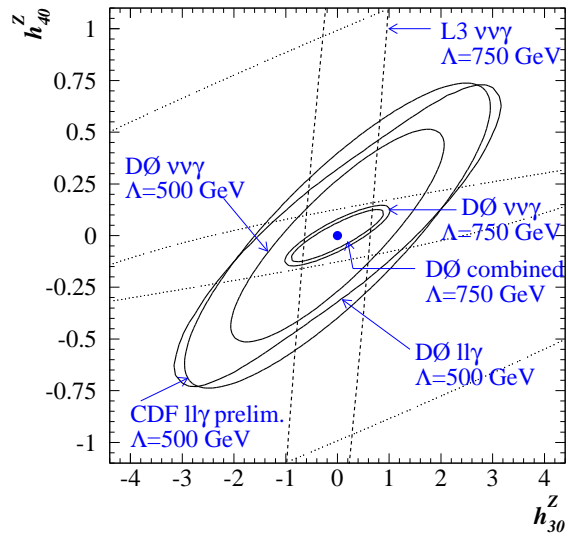}}
\caption{
{\footnotesize
Summary of the 95\% CL limits on anomalous CP-conserving \zzg\
coupling parameters. The CDF and \d0\ limits are indicated by the
\textit{solid contours} and the \textit{dashed contour} indicates the limits from 
L3~\cite{L3} for
$\Lambda_{FF}=500$~GeV. The \textit{dotted contours} show the unitarity limits for
$\Lambda_{FF}=500$~GeV (\textit{outer contour}) and $\Lambda_{FF}=750$~GeV (\textit{inner
contour}).}
}
\label{fig:zg_limits_sum}
\end{figure}
\begin{figure}[]
    \epsfysize = 9cm
    \centerline{\epsffile{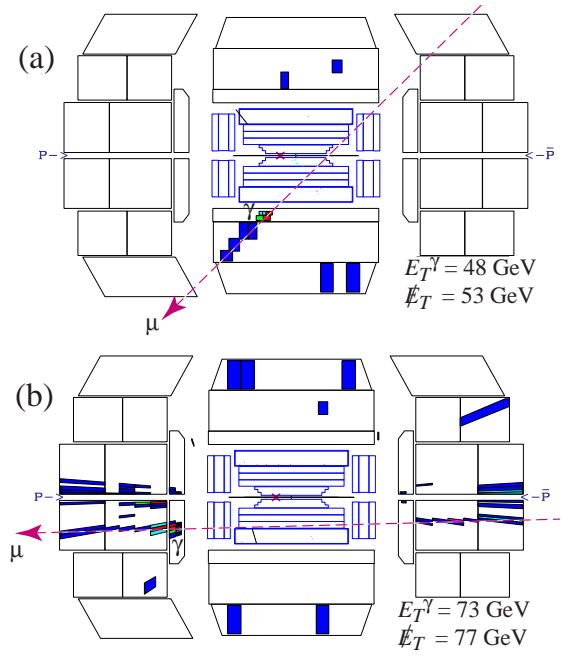}}
\caption{
{\footnotesize
Side views of the \d0\ calorimeter and
tracking systems for events attributed to (\textit{a}) cosmic ray muon
bremsstrahlung and (\textit{b}) beam halo muon bremsstrahlung. The \textit{dashed line}
indicates the probable muon track, the photon is labeled
$\gamma$, and the reconstructed event vertex is indicated by the \textit{cross}.
All cells in
the calorimeter with energy greater than 200~MeV are shown.}
}
\label{fig:muon_brem}
\end{figure}
\begin{figure}[p]
    \epsfysize = 3.5cm
    \centerline{\epsffile{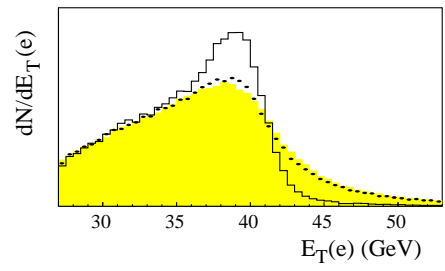}}
\vskip -2mm
\caption{
{\footnotesize
The $E_T^e$ spectrum for $W \to e \nu$ events with $p_T^W = 0$
(\textit{solid line}), with the correct $p_T^W$ distribution ($\bullet$) and
with the \d0\ detector resolutions (\textit{shaded}) [from~\cite{d0_wmass1b_prd}]}
}
\label{fig:jacobian}
\end{figure}

\bigskip
{\footnotesize
\begin{thebibliography}{99}
%
%

\def\order#1#2#3{#1:#2 (#3)}

\def\app#1#2#3{{\it Acta Phys. Pol. {B}}\order{#1}{#2}{#3}}
\def\arnps#1#2#3{{\it Ann. Rev. Nucl. Part. Sci. }\order{#1}{#2}{#3}}
%
\def\ijmpa#1#2#3{{\it Int. J. Mod. Phys. {A}}\order{#1}{#2}{#3}}
%
\def\jpa#1#2#3{{\it J. Phys. {A}}\order{#1}{#2}{#3}}
%
\def\nci#1#2#3{{\it Nuov. Cim }\order{#1}{#2}{#3}}
\def\nima#1#2#3{{\it Nucl. Instrum. Methods {A}}\order{#1}{#2}{#3}}
\def\nimb#1#2#3{{\it Nucl. Instrum. Methods {B}}\order{#1}{#2}{#3}}
\def\npa#1#2#3{{\it Nucl. Phys. {A}}\order{#1}{#2}{#3}}
\def\npb#1#2#3{{\it Nucl. Phys. {B}}\order{#1}{#2}{#3}}
%
\def\physa#1#2#3{{\it Physica {A }}\order{#1}{#2}{#3}}
\def\physd#1#2#3{{\it Physica {D }}\order{#1}{#2}{#3}}
\def\plb#1#2#3{{\it Phys. Lett. {B }}\order{#1}{#2}{#3}}
\def\pl#1#2#3{{\it Phys. Lett.\ }\order{#1}{#2}{#3}}
\def\ppnp#1#2#3{{\it Prog. Part. Nucl. Phys.\ }\order{#1}{#2}{#3}}
\def\pr#1#2#3{{\it Phys. Rev.\ }\order{#1}{#2}{#3}}
\def\prep#1#2#3{{\it Phys. Rep.\ }\order{#1}{#2}{#3}}
\def\prl#1#2#3{{\it Phys. Rev. Lett.\ }\order{#1}{#2}{#3}}
\def\pra#1#2#3{{\it Phys. Rev. {A }}\order{#1}{#2}{#3}}
\def\prd#1#2#3{{\it Phys. Rev. {D }}\order{#1}{#2}{#3}}
\def\ptp#1#2#3{{\it Prog. Theo. Phys.\ }\order{#1}{#2}{#3}}
%
\def\itp#1#2#3{{\it Rev. Mod. Phys.\ }\order{#1}{#2}{#3}}
\def\rpp#1#2#3{{\it Rep. Progr. Phys.\ }\order{#1}{#2}{#3}}
%
\def\solp#1#2#3{{\it Solar Phys.}\order{#1}{#2}{#3}}
\def\sjnp#1#2#3{{\it Soviet J. Nucl. Phys.}\order{#1}{#2}{#3}}
%
\def\yadfiz#1#2#3{{\it Yad. Fiz. (Phys. At. Nucl.)}\order{#1}{#2}{#3}}
%
\def\zphysold#1#2#3{{\it Z. Phys. }\order{#1}{#2}{#3}}
\def\zphys#1#2#3{{\it Z. Phys. {C }}\order{#1}{#2}{#3}}

\bibitem{sm.general}
Weinberg S. \prl{19}{1264}{1967}.
Salam A. In {\it Elementary Particle Theory}, ed. N Svarthom, et al. 
Stockholm: (1968);
Peskin ME, Schroeder DV. {\it An Introduction to Quantum Field Theory}. Addison-Wesley (1995);
Weinberg S. {\it The Quantum Theory of Fields}. Cambridge, UK: Cambridge Univ. Press (1995)
%
\bibitem{tbc.general}
Baur U, Errede S, Muller T, eds. {\it Proc. Int. Symp. Vector Boson Self-interact.} {\it Am. Inst. Phys.} (1996) (AIP Conf. Proc. 350)
%
\bibitem{leff.general}
Weinberg S. \physa{96}{327}{1979};
Georgi H. \npb{361}{339}{1991}, \npb{363}{301}{1991};
Pich A. In {\it Mex. Sch. Part. Fields, 5th}, Guanajuato, Mex., Nov. 30--Dec 11, 1992, ed. JL Lucio, M Vargas. Am. Inst. Phys. (1994) (AIP Conf. Proc., v. 317)
%
\bibitem{new}
Weinberg S. In {\it Conference on Historical and Philosophical
Reflections on the Foundation of Quantum Field Theory}, Boston, MA, Mar. 1--3, 1996 (e-print archive: hep-th/9702027)
%
\bibitem{models}
{\it Proc. Int. Europhys. Conf. High Energy Phys.}, Aug. 19--26, 1997, Jerusalem, Israel;
{\it Proc. 1996 DPF/DPB Summer Stud. New Dir. High-Energy Phys. (Snowmass 96)}, Snowmass, CO, Jun. 25--Jul. 12, 1996, ed. DG Cassel, L Trindle Gennari, RH Siemann (Stanford Linear Accelerator Center, 1997); {\it Proc. Int. Conf. Phys. Beyond Standard Model, 5th}, Balholm, Norway, Apr. 29--May 4, 1997
%
\bibitem{pdg}
Barnett RM, et al. \prd{54}{1}{1996}
%
\bibitem{chiral.leff}
Weinberg S. \physa{96}{327}{1979}; Chanowitz MS, Gaillard MK. \npb{261}{379}{1985};
Chanowitz M, Golden M, Georgi H. \prl{57}{2344}{1986}
%
\bibitem{frere.et.al.}
Frere JM, et al.  \plb{292}{348}{1992}; \npb{429}{3}{1994}
%
\bibitem{perez.et.al.}
Perez MA, Toscano JJ, Wudka J. \prd{52}{494}{1995}
%
\bibitem{veltman.ir.uv.}
Veltman M. \app{12}{437}{1981}
%
\bibitem{arzt.et.al.mom}
 Arzt C, Einhorn MB, Wudka J, \prd{49}{1370}{1994}
%
\bibitem{eom} 
Georgi H. \npb{361}{339}{1991}, \npb{363}{301}{1991};
De Rujula A, et al. \npb{384}{3}{1992}; 
Arzt C. \plb{342}{189}{1995}
%
\bibitem{bw}
Buchmuller W, Wyler D, \npb{268}{621}{1986}
%
\bibitem{hagiwara.peccei}
Hagiwara K, et al. {\it Nucl. Phys.} B282:253 (1987)
%
\bibitem{renard-moments}
Renard FM, {\it Nucl. Phys.} B196:93 (1982)
%
\bibitem{future.39}
Gounaris G, et al. ``Triple Gauge Boson Couplings'' in {\it
Proceedings of the CERN Workshop on LEP~II Physics}, ed. Altarelli G
et al, CERN 96-01, Vol. 1, p.525 (1996)
%
\bibitem{appelquist.wu}
Longhitano AC. \prd{22}{1166}{1980}; \npb{188}{118}{1981};
Holdom B. \plb{258}{156}{1991};
Appelquist T, Wu G-H. \prd{48}{3235}{1993}; \prd{51}{240}{1995}
%
\bibitem{hisz.ref}
Hagiwara K, et al. \prd{48}{2182}{1993}
%
\bibitem{arzt.et.al.loops}
Arzt C, Einhorn MB, Wudka J. \npb{433}{41}{1995}
%
\bibitem{wudka.rev}
Wudka J. \ijmpa{9}{2301}{1994}
%
\bibitem{nda}
Weinberg S. \physa{96}{327}{1979};
Georgi H, Manohar A. \npb{234}{189}{1984};
Georgi H. \plb{298}{187}{1993}
%
\bibitem{polchinski}
Polchinski J. {\it Recent Directions in Particle Theory: From Superstrings and Black
   Holes to the Standard Model (TASI 92)} Boulder, CO, Jun 3--28, 1992, ed. J Harvey, J Polchinski. World Scientific (1993) 
%
\bibitem{unit.form.factors}
Baur U, Berger EL. \prd{47}{4889}{1993};
Golden M, Han T, Valencia G. In {\it Electroweak Symmetry Breaking 
and New Physics at the TeV Scale}, ed. TL Barklow, et al. World
Scientific, Singapore (1996)
%
\bibitem{fer}
For similar calcualtions see
A. Culatti, \etal, in {\it  Physics with $e^+ e^-$  
Linear Colliders} (The European Working Groups 4 Feb - 1 Sep 1995:  
Session 1), Annecy, France, 4 Feb 1995, and to  
{\it 3rd Workshop on Physics and Experiments with $e^+ e^-$ Linear Colliders  
(LCWS 95)}, Iwate, Japan, 8-12 Sep 1995. 
F. Feruglio and S. Rigolin, \plb{397}{245}{1997}
%
\bibitem{bound.60}
Grossman Y, Ligeti Z, Nardi E. \npb{465}{369}{1996} \npb{480}{753}{1996}
%
\bibitem{bound.23}
Alam S, Dawson S, Szalapski R. \prd{57}{1577}{1998}
%
\bibitem{bound.28.bound.29.bound.48}
Grau A, Grifols JA. \plb{154}{283}{1985};
Spagnolo S. Report INFN-TH-97-01 (unpublished)
Herzog F. \plb{148}{355}{1984}, \plb{155}{468}{1985}
%
\bibitem{zgv.29}
Baur U, Berger EL. \prd{47}{4889}{1993}
%
\bibitem{wwv.5prime}
Boudjema F, et al. \prd{43}{2223}{1991}
%
\bibitem{bound.54.bound.27.bound.1}
Marciano WJ, Queijeiro A. \prd{33}{3449}{1986};
De Rujula A, et al.  \npb{384}{3}{1992}
He XG, Ma JP, McKellar BHJ. \plb{304}{285}{1993}
%
\bibitem{bound8.bound10.bound49.bound59}
Numata K. \zphys{52}{691}{1991}; 
Baur U. In {\it Workshop B Phys. Hadron Accel.}, Snowmass, CO, Jun. 21--Jul. 2, 1993, ed. CS Mishra,  P McBride. Fermilab (1994) (SSCL-SR-1225,C93/06/21);
Hewett JL. In {\it Spin Structure in High Energy Processes}, ed. L DePorcel, C Dunwoodie. SLAC (1994) (SLAC-Report-444,C93/07/26)
Sinha N, Sinha R. Report IMSC-97-03-08 (unpublished) (e-print archive: hep-ph/9707416)
%
\bibitem{bound.55}
Godfrey S, Konig H. \prd{45}{3196}{1992}
%
\bibitem{pinch}
Cornwall JM. In {\it  Fr.-Am. Sem. Theor. Aspects Quantum Chromodynamics}, Marseille, France, Jun. 9--13, 1981, ed. JW Dash. (CPT-81/P-1345,C81/06/09.1); \prd{26}{1453}{1982};
Argyres EN, et al. \npb{391}{23}{1993};
Papavassiliou J, Philippides K. \prd{48}{4255}{1993}
%
\bibitem{sm19}
Papavassiliou J, Philippides K. \prd{48}{4255}{1993}
%
\bibitem{sm6}
Kodaira J, et al. In {\it INS Workshop Phys. $e^+ e^-$, $ e^- \gamma $ and $ \gamma \gamma $ Collis. at Linear Accel.}, Tokyo, Japan, Dec. 20--22, 1994, ed. Z Hioki, T Ishii, R Najima. Tokyo Univ., Inst. Nucl. Study (1995) 
%
\bibitem{sm13}
Chang D, Keung W-Y, \npb{355}{295}{1991} 
%
\bibitem{sm20}
Khriplovich IB, Pospelov ME. \npb{420}{505}{1994}
%
\bibitem{2hdm5}
Couture G, et al. \prd{36}{859}{1987} 
%
\bibitem{sm12}
He X-G, McKellar BHJ. \prd{42}{3221}{1990}, \prd{50}{4719}{1994}
%
\bibitem{sm11}
Couture G, Ng JN \zphys{35}{65}{1987}
%
\bibitem{susy2}
Argyres EN, et al. \plb{383}{63}{1996} 
%
\bibitem{susy7}
Kadoyoshi T, Oshimo N. \prd{55}{1481}{1997}
%
\bibitem{tc2}
Appelquist T, Wu G-H. \prd{48}{3235}{1993}; \prd{51}{240}{1995}
%
\bibitem{various10}
Burgess CP, Pilaftsis A. \plb{333}{427}{1994}
%
\bibitem{chang.et.al.}
Chang D, Keung W-Y, Pal PB. \prd{51}{1326}{1995}
%
\bibitem{barroso}
Barroso A, Boudjema F, Cole J, Dombey N. \zphys{28}{149}{1985}

%
%
\bibitem{wg-lit}
Mikaelian KO, {\it Phys. Rev.} D17:750 (1978);
Brown RW, Sahdev D, Mikaelian KO, {\it Phys. Rev.} D20:1164 (1979);
Mikaelian KO, Samuel MA, Sahdev D, {\it Phys. Rev. Lett.} 43:746 (1979);
Zhu Dongpei, {\it Phys. Rev.} D22:2266 (1980);
Goebel CJ, Halzen F, Leveille JP, {\it Phys. Rev.} D23:2682 (1981);
Brodsky SJ, Brown RW, {\it Phys. Rev. Lett.} 49:966 (1982);
Smauel MA, {\it Phys. Rev.} D27:2724 (1983);
Brown RW, Kowalski KL, Brodsky SJ, {\it Phys. Rev.} D28:624 (1983);
Bilchak CL, Brown RW, Stroughair JD, {\it Phys. Rev.} D29:375 (1984);
Cortes J, Hagiwara K, Herzog F, {\it Nucl. Phys.} B278:26 (1986);
Wallet JC, {\it Z. Phys.} C30:575 (1986)
\bibitem{Baur&Zeppenfeld-wgmc}
Baur U, Zeppenfeld D. {\it Nucl. Phys.} B308:127 (1988)
\bibitem{Baur&Berger-wgmc}
Baur U, Berger EL. {\it Phys. Rev.} D41:1476 (1990)
\bibitem{zg-nlo}
Baur~U, Han~T, Ohnemus J.  hep-ph/9710416 (1997)
\bibitem{Baur&Berger-zgmc}
Baur U, Berger EL. {\it Phys. Rev.} D47:4889 (1993)
\bibitem{Baur&Errede&Ohnemus}
Baur U, Errede S, Ohnemus J. {\it Phys. Rev.} D48:4103 (1993)
\bibitem{wwxs_lo}
Hagiwara K, Woodside J, Zeppenfeld D.  {\it Phys. Rev.} D41:2113 (1990)
\bibitem{wgzgxs_nlo}
Ohnemus J.  {\it Phys. Rev.} D47:940 (1993)
\bibitem{wwxs_nlo}
Ohnemus J.  {\it Phys. Rev.} D44:1403 (1991)
\bibitem{wzxs_nlo}
Ohnemus J.  {\it Phys. Rev.} D44:3477 (1991)
\bibitem{wg-nlo}
Baur U, Han T, Ohnemus J. {\it Phys. Rev.} D48:5140 (1993)
\bibitem{ww-nlo}
Baur U, Han T, Ohnemus J. {\it Phys. Rev.} D53:1098 (1996)
\bibitem{wz-nlo}
Baur U, Han T, Ohnemus J. {\it Phys. Rev.} D51:3381 (1995)
\bibitem{baur-unitarity}
Baur U, Zeppenfeld D. {\it Phys. Lett.} B201:383 (1988)
\bibitem{cdf_wgzg1a_prd}
Abe~F, et al (CDF Collaboration). Fermilab-Pub-94/244-E (1994)
%
%
\bibitem{accelerators}
Thompson~J. Fermilab TM-1909 (1994)
\bibitem{Tevatron}
Edwards~H. {\it {Ann. Rev. Nucl. Part. Sci.}} 35:605 (1985)
\bibitem{pbar-source}
Church~MD, Marriner~JP. {\it {Ann. Rev. Nucl. Part. Sci.}} 43:253 (1993)
\bibitem{CDF-detector}
Abe~F, et al (CDF Collaboration). {\it Nucl. Instr. and Meth.} A271:387 (1988)
\bibitem{D0-detector}
Abachi~S, et al (\d0\ Collaboration). {\it Nucl. Instr. and Meth.}
A338:185 (1994)
\bibitem{cdf_wwlvjj1b}
Nodulman~LJ. In {\it Int. Conf. High Energy Phys., 28th}, Warsaw, Poland (1996)
\bibitem{CDF-topPRD}
Abe~F et al (CDF Collaboration). {\it Phys. Rev.} D50:2966 (1994)
%
%
%
%
\bibitem{cdf_wg1a_prl}
Abe~F, et al (CDF Collaboration). {\it Phys. Rev. Lett.} 74:1936 (1995)
\bibitem{d0_wg1a_prl}
Abachi~S, et al (\d0\ Collaboration). {\it Phys. Rev. Lett.} 75:1034 (1995) 
\bibitem{d0_diboson_1a_prd}
Abachi~S, et al (\d0\ Collaboration). {\it Phys. Rev.} D56:6742 (1997)
\bibitem{d0_wg1b_prl}
Abachi~S, et al (\d0\ Collaboration). {\it Phys. Rev. Lett.} 78:3634
(1997)
\bibitem{cdf_wg1b}
Benjamin~D. In {\it ``Topical Workshop Proton-Antiproton
Collider Phys., 10th},'' Batavia, IL. AIP Conference Proceedings 357,
ed. Raja R, Yoh J, p.370. 
 (1996) 
\bibitem{cousins}
Cousins~D, Highland~V. {\it Nucl. Instr. Meth.} A320:331 (1992)
\bibitem{landsberg-thesis}
Landsberg GL. Ph.D. Thesis, State Univ. NY Stony
Brook (1994)
%
%
\bibitem{bsgamma}
Drell PS. In {\it ``Int. Symp. Lepton-Photon
Interact., XVIIIth''}, Hamburg, Germany (1997) 
\bibitem{d0_1a_wwdilep}
Abachi~S, et al (\d0\ Collaboration). {\it Phys. Rev. Lett.} 75:1023
(1995)
\bibitem{cdf_wwdilep}
Abe~F, et al (CDF Collaboration). {\it Phys. Rev. Lett.} 78:4536 (1997)
\bibitem{d0_1b_wwdilep}
Yasuda T. In {\it ``Hadron Collider Physics XII''}, Stony Brook,
NY (1997)
\bibitem{cdf_wwlvjj1a_prl}
Abe~F, et al (CDF Collaboration). {\it Phys. Rev. Lett.} 75:1017 (1995)
\bibitem{d0_wwlvjj1a_prl}
Abachi~S et al (\d0\ Collaboration). {\it Phys. Rev. Lett.} 77:3303
(1996)
\bibitem{d0_wwlvjj1b_prl}
Abachi~S, et al (\d0\ Collaboration). {\it Phys. Rev. Lett.} 79:1441
(1997)
\bibitem{vecbos}
Berends FA, et al. {\it Nucl. Phys.} B357:32 (1991)
\bibitem{herwig}
Marchesini G, et al. {\it Comput. Phys. Commun.} 67:465 (1992)
\bibitem{geant}
Brun~R, et al. \geant\ User's Guide v3.14, CERN Program Libr.
\bibitem{lep-tgc-results}
Ward~D. In {\it ``Int. Europhys. Conf. High Energy 
Phys.},'' Jerusalem, Israel (1997)
\bibitem{cdf_zg1a_prl}
Abe~F, et al (CDF Collaboration). {\it Phys. Rev. Lett.} 74:1941 (1995)
\bibitem{d0_zg1a_prl}
Abachi~S, et al (\d0\ Collaboration). {\it Phys. Rev. Lett.} 75:1028
(1995)
\bibitem{d0_zg1b_prl}
Abbott~B, et al (\d0\ Collaboration). Fermilab-Pub-97/363-E (1997). {\it Phys. Rev.} D In press
\bibitem{L3}
Acciarri M, et al (L3 Collaboration). {\it Phys. Lett.} B412:201 (1997)
\bibitem{d0_zvvg_prl}
Abachi~S, et al (\d0\ Collaboration). {\it Phys. Rev. Lett.} 78:3640
(1997)
\bibitem{d0_wmass1b_prd}
Abbott~B, et al (\d0\ Collaboration). Fermilab-Pub-97/422-E (1997). {\it Phys. Rev.} D In press
%
%
\bibitem{tev2000}
Amidei D, Brock R, eds. Fermilab-Pub-96/082 (1996)
\bibitem{LEPII-yellow}
Ajaltouni Z, et al. In {\it Physics at LEP2}, ed. G Altarelli,
T Sj\"{o}strand. CERN 96-01 (1996)
%
\bibitem{lhc.general}
Womersley J. In {\it Int. Workshop Fundam. Probl. High-Energy Phys. Field Theory, 20th}, Protvino, Russia, June 24--26, 1997
Hinchliffe I. In {\it ITP Conf. Future High Energy Colliders}, Santa Barbara, CA, Oct. 21--25, 1996, ed. Z Parsa (Am. Inst. Phys., 1997) (AIP Conf. Proc., 397)
%
\bibitem{nlc.general} 
Kuhlman S, et al (NLC ZDR Design Group and NLC Phys. Working Group). Rep. SLAC-R-0485 (e-print archive: hep-ex/9605011)
%
\bibitem{wwv.4}
Baur U, Han T, Ohnemus J. Rep. UB-HET-97-03 (unpublished) (e-print archive: hep-ph/9710416)
%
\bibitem{abraham}
Abraham KJ, Lampe B. Rep. MPI-PHT-98-10 (unpublished) (e-print archive: hep-ph/9801400)
%
\bibitem{future.10}
Hinchliffe I, Womersley J. In {\it 1996 DPF/DPB Summer Stud. New Dir. High-Energy Phys. (Snowmass 96)}, Snowmass, CO, Jun. 25--Jul. 12 1996, ed. DG Cassel, L Trindle Gennari, RH Siemann (Stanford Linear Accelerator Center, 1997)
%
\bibitem{atlas}
Fouchez D. ATLAS internal note PHYS-NO-160 (1994)
%
\bibitem{future.11}
Baur U, Han T, Ohnemus J. \prd{53}{1098}{1996}
%
\bibitem{zgv.2}
Baur U, Han T, Ohnemus J. Rep. UB-HET-97-03 (unpublished) (e-print archive: hep-ph/9710416)
%
\bibitem{future.26}
Baur U, et al. In {\it Large Hadron Collider Workshop (LHC)}, Aachen,
Germany, Oct. 4--9, 1990, ed. G Jarlskog, D Rein, p.570, CERN 90-10 (1990)
%
\bibitem{ginzburg}
Ginzburg IF, et al. \nima{l219}{5}{1984}
Telnov VI. \nima{294}{72}{1990}
%
\bibitem{wwv.3.wwv.5.wwv.13.wwv.15.wwv.16}
Likhoded AA, Valencia G, Yushchenko OP. Rep. BINP-97-11 (unpublished) (e-print archive: hep-ph/9711325)
Boudjema F et al. \prd{43}{2223}{1991}
de Campos F, et al. \prd{56}{4384}{1997}
Boudjema F. In {\it Workshop Phys. Exper. Linear Colliders}, Morioka-APPI, Japan, Sep. 8--12, 1995, ed. A Miyamoto, et al. World Sci. (1996)
Baillargeon M, Belanger G, Boudjema F. \npb{500}{224}{1997}
%
\bibitem{wwv.6}
Pankov AA, Paver N. In {\it Workshop $e^+ e^- $ Collis. TeV Energies: Phys. Potential}, Annecy, France, Feb. 4, 1995: Session 1 (Session 2: Jul. 2--3, 1995, Assergi, Italy; Session 3: Aug. 30--Sep. 1, 1995, Hamburg, Germany),  ed. PM Zerwas. DESY (1996)
%
\bibitem{wwv.12}
Rindani SD, Singh JP. Report FTUV-97-12 (unpublished) (e-print archive: hep-ph/9703380)
%
\bibitem{zgv.11.zgv.4}
Choudhury D, Rindani SJ. \plb{335}{198}{1994}; 
Rizzo TG. In {\it 1996 DPF/DPB Summer Stud. New Dir. High-Energy Phys. (Snowmass 96)}, Snowmass, CO, Jun 25--Jul. 12, 1996, ed. DG Cassel, L Trindle Gennari, RH Siemann. SLAC (1997)



 
\end{thebibliography}
}
\end{document}